\documentclass[11pt]{article}

\usepackage{html,amsmath,amssymb,amsfonts,epsfig,cancel,cite,longtable,caption,siunitx,array,color,booktabs,setspace,mathtools}

\usepackage{dsfont}

\newlength{\xtrawidth}
\newlength{\xtraheight}
\renewcommand\baselinestretch{1.23}
\numberwithin{equation}{section}

\newcommand{\be}{\begin{equation}}
\newcommand{\ee}{\end{equation}}
\newcommand{\beq}{\begin{equation}}
\newcommand{\eeq}{\end{equation}}
\newcommand{\ba}{\begin{array}}
\newcommand{\ea}{\end{array}}
\newcommand{\bea}{\begin{eqnarray}}
\newcommand{\eea}{\end{eqnarray}}
\newcommand{\bean}{\begin{eqnarray*}}
\newcommand{\eean}{\end{eqnarray*}}

\newcommand{\tr}[1]{\textcolor{red}{#1}}

\newcommand{\IC}{\mathbb{C}}
\newcommand{\IP}{\mathbb{P}}

\newcommand{\IZ}{\mathbb{Z}}
\newcommand{\IQ}{\mathbb{Q}}
\newcommand{\cO}{{\cal O}}
\newcommand{\cN}{{\cal N}}

\newcommand{\cA}{{\cal A}}
\newcommand{\cB}{{\cal B}}
\newcommand{\cC}{{\cal C}}

\newcommand{\cV}{{\cal V}}

\newcommand{\fn}{\footnotesize}

\def\cjn1{{\cA, \cC^*\otimes \wedge^j \cN^*}}
\def\bjn1{{\cA, \cB^*\otimes \wedge^j \cN^*}}
\def\vjn1{{\cA, \cV^*\otimes \wedge^j \cN^*}}
\def\cjn2{{\cA, \cC\otimes \wedge^j \cN^*}}
\def\bjn2{{\cA, \cB\otimes \wedge^j \cN^*}}
\def\vjn2{{\cA, \cV\otimes \wedge^j \cN^*}}

\newcommand{\cicy}[2]{\begin{matrix} #1\end{matrix}\!\left[\begin{matrix}#2 \end{matrix}\right]}

\newcommand{\capt}[3]{\parbox{#1}{\renewcommand{\baselinestretch}{1.0}
                                                           \caption{\label{#2}\small\it #3}}}
\newcommand{\lrarr}{\longrightarrow }

\def\str{\vrule height7pt depth5pt width0pt}
\def\strr{\vrule height8pt depth5pt width0pt}

\definecolor{db}{rgb}{0,0,0.8}
\definecolor{dg}{rgb}{.1,0.5,.5}
\definecolor{dr}{rgb}{0.9,0,0}

\begin{document}

\title{{\LARGE \bf Hodge Numbers for All CICY Quotients \\[4mm]}}

\author{
Andrei Constantin$^{1}$,
James Gray$^2$, 
Andre Lukas$^3$ 
}
\date{}
\maketitle
\thispagestyle{empty}
\begin{center} { 
${}^1\,$\itshape Department of Physics and Astronomy, Uppsala University, \\ 
       SE-751 20, Uppsala, Sweden\\[0.3cm]
${}^2\,$Physics Department, Robeson Hall, Virginia Tech,\\ Blacksburg, VA 24061, USA\\[0.3cm]
${}^3\,$Rudolf Peierls Centre for Theoretical Physics, Oxford University,\\
       1 Keble Road, Oxford, OX1 3NP, U.K.}\\
       
\end{center}

\vspace{1cm}
\abstract
\noindent 
We present a general method for computing Hodge numbers for Calabi-Yau manifolds realised as discrete quotients of complete intersections in products of projective spaces. The method relies on the computation of equivariant cohomologies and is illustrated for several explicit examples. In this way, we compute the Hodge numbers for all discrete quotients obtained in Braun's classification \cite{Braun:2010vc}.

\vskip 7.0cm
{\hbox to 7cm{\hrulefill}}
\noindent
{\fn andrei.constantin@physics.uu.se\\ jamesgray@vt.edu\\ lukas@physics.ox.ac.uk} 
\newpage

%\tableofcontents

%
%==========
%

\section{Introduction}
Complete intersection Calabi-Yau manifolds in products of projective spaces (or CICY manifolds, for short) were first constructed by Yau in Ref.~\cite{Yau:1986gu} and then H\"ubsch and Green and H\"ubsch in Refs.~\cite{Hubsch:1986ny,Green:1986ck}. Ever since then, they have provided a fruitful arena for studying string compactifications. The construction was employed in Refs.~\cite{Candelas:1987kf, Candelas:1987du} in order to compile an exhaustive list of complete intersection Calabi-Yau three-folds and in Refs.~\cite{Gray:2013mja, Gray:2014fla} for the case of four-folds. The CICY construction was recently generalised in Ref.~\cite{Anderson:2015iia} to include manifolds defined using local sections of mixed degree line bundles over products of projective spaces. 

Over the years, CICY manifolds have been employed in various string compactifications, including compactifications of the heterotic string (see, for example,~Refs.~\cite{Braun:2005ux,Braun:2005bw, Bouchard:2005ag, Braun:2009qy, Braun:2011ni, Anderson:2007nc, Anderson:2008uw, Anderson:2009sw, Anderson:2009nt, Anderson:2010tc,Anderson:2010mh, Anderson:2010ty,Anderson:2011cza, Anderson:2011ty, Anderson:2013qca, Anderson:2009ge, Anderson:2009mh, Anderson:2011ns,  Anderson:2012yf, Anderson:2013xka, Buchbinder:2013dna, Buchbinder:2014sya, Buchbinder:2014qda, Anderson:2014hia, Buchbinder:2014qca, Nibbelink:2015ixa, Nibbelink:2015vha, Constantin:2015bea, Blesneag:2015pvz, Buchbinder:2016jqr} for some recent developments). For model building purposes, particularly in the context of heterotic theory, the existence of freely acting discrete symmetry groups on CICY manifolds plays a crucial role. Dividing the original CICY by a freely-acting symmetry produces a non-simply connected Calabi-Yau manifold, which in turn can be decorated with discrete Wilson lines that break the intermediate GUT group to the Standard Model group. The possibility of constructing realistic models from the $E_8\times E_8$ heterotic string without an intermediate GUT phase, originally studied in \cite{Blumenhagen:2005ga,Blumenhagen:2006ux,Blumenhagen:2006wj}, has been recently investigated in \cite{Anderson:2014hia}, showing that there is a tension between directly breaking $E_8$ to the Standard Model group and obtaining a particle spectrum free from exotics. This suggests that non-simply connected Calabi-Yau manifolds are indeed a crucial ingredient for realistic model building in the context of the heterotic string. More generally, finding the Hodge numbers for quotient Calabi-Yau manifolds is an important task for string model building. This information has already been used in the large scale searches \cite{Anderson:2009mh,Anderson:2011ns, Anderson:2012yf, Anderson:2013xka} for heterotic vacua with a realistic particle physics spectrum. The Hodge numbers are also important to determine the size of the K\"ahler and complex structure moduli space and, therefore, enter the discussion of moduli stabilization\cite{Anderson:2009sw, Anderson:2009nt, Anderson:2010mh, Anderson:2011cza, Anderson:2011ty, Anderson:2013qca}. 

A systematic study of discrete symmetry groups, $G$, on CICY manifolds $X$ was initiated by Candelas and Davies in Ref.~\cite{Candelas:2008wb}, and completed by Braun in Ref.~\cite{Braun:2010vc}, who classified, through an automated scan, all finite group actions that descend from linear automorphisms of the ambient space, given by a product of projective spaces, to free actions on the CICY manifold. Clearly, Braun's classification depends on the particular embedding of the Calabi-Yau manifold into the ambient space, as constructed in Ref.~\cite{Candelas:1987kf}.  Since this embedding is not unique, it is expected that other discrete symmetries can be found by considering different embeddings of the same manifolds. Be that as it may, Braun's classification reveals that $195$ out of the  $7890$ manifolds in the CICY list admit freely acting discrete symmetry groups. Many of these $195$ CICY manifolds admit multiple freely-acting discrete symmetries for a total of $1695$ symmetries on these manifolds. 

The enterprise of systematically computing Hodge numbers for CICY quotients $X/G$ has been undertaken in Refs.~\cite{Candelas:2008wb, Candelas:2010ve, Candelas:2015amz} and summarised in Ref.~\cite{Candelas:2016fdy}, using methods such as the counting of parameters in the defining polynomials, the counting of K\"ahler classes for favourable embeddings (including favourable embeddings in products of spaces containing factors of del Pezzo surfaces). Though fruitful in many cases, the above methods were not applicable to a significant number of quotients ($\sim$ 300 quotients, mostly $\IZ_2$-quotients, on $\sim$ 70 manifolds were left out). The purpose of the present note is to establish a generic algorithm, relying on the computation of equivariant cohomologies, which can be applied in an exhaustive manner to the computation of Hodge numbers for all CICY quotients. Thus we aim to complete this task for the manifolds that were missed in Refs.~\cite{Candelas:2008wb, Candelas:2010ve, Candelas:2015amz} and at the same time provide an independent check of the previous results. It is remarkable that our results agree with those found in the above references, in all $1426$ cases that we could compare. In the implementation of the algorithm described below we have made use of the CICY package \cite{cicypackage}, especially for the computation of (equivariant) line bundle cohomologies.

In the next section, we begin by explaining how the cohomology $H^{2,1}(X)$ can be computed using the normal bundle sequence and the Euler sequence. Further, we show how this computation can be carried over to the quotient manifold $X/G$ by writing down the equivariant structures of the various bundles involved. In Section 3 we illustrate this method for several explicit examples. We conclude in Section 4 by providing the updated Hodge number plot in Figure~\ref{TipHodgePlot}. The detailed results of our computation are given in Appendix~\ref{appendix}, where we have tabulated the Hodge numbers of all CICY quotients.

\section{General method}
In this section, we explain the general method to compute Hodge numbers of CICY quotients. We being with a general set-up of CICY manifolds $X$ and first review how to compute the ``upstairs" cohomology $H^{2,1}(X)$ of these manifolds. Then we introduce freely-acting discrete symmetries $G$ and consider the quotient $\tilde{X}=X/G$. The ``downstairs" cohomology $H^{2,1}(\tilde{X})$ of this quotient is given by the $G$-invariant part of the upstairs cohomology $H^{2,1}(X)$ and we explain in detail how to calculate this $G$-invariant part. In this way, we can obtain the Hodge number $h^{2,1}(\tilde{X})$ of the quotient. Since the Euler number of the quotient is easily obtained from its upstairs counterpart by dividing by the group order $|G|$, this fixes $h^{1,1}(\tilde{X})$ as well.

\subsection{CICY manifolds and upstairs cohomology}
The CICY manifolds are embedded in ambient spaces of the form ${\cal A}=\mathbb{P}^{n_1}{\times}{\ldots}{\times}\, \IP^{n_m}$, consisting of $m$ projective factors with dimensions $n_r$, where $r=1,\ldots ,m$. The homogeneous ambient space coordinates for each $\mathbb{P}^{n_r}$ factor are denoted by ${\bf x}_r=(x_r^0,\ldots ,x_r^{n_r})$ and, collectively for all of ${\cal A}$, by ${\bf x}=({\bf x}_1,\ldots ,{\bf x}_m)$. The CICY three-folds $X\subset \cA$ are defined as the common zero locus of $K$ polynomials, $p_a$, where $a=1,\ldots, K$, each with multi-degree ${\bf q}_a=(q_a^1,\ldots ,q_a^m)$. This means that the polynomial $p_a$ has degree $q_a^r$ in the homogeneous coordinates ${\bf x}_r$ of the $r^{\rm th}$ projective factor of the ambient space. This information is frequently summarised by writing down the configuration matrix
\begin{equation}
 X=\left[\begin{array}{c|ccc}\mathbb{P}^{n_1}&q_1^1&\cdots&q_K^1\\
                                           \vdots&\vdots&\vdots&\vdots\\
                                            \mathbb{P}^{n_m}&q_1^m&\cdots&q_K^m\end{array}\right]^{h^{1,1}(X),h^{2,1}(X)}_{\eta(X)}\; , \label{Q}
\end{equation}                                            
where the Hodge numbers $h^{1,1}(X)$, $h^{2,1}(X)$ are attached as a superscript and the Euler number $\eta(X)$ as a subscript. In order for this data to define a Calabi-Yau three-fold we require that
\begin{equation}
 \sum_{r=1}^mn_r-K=3\;,\qquad\qquad \sum_{a=1}^Kq_a^r=n_r+1\quad\mbox{for all}\quad r=1,\ldots m\; .
\end{equation} 
The first of these equations ensures that the manifold $X$ is indeed complex three-dimensional, while the second equation is equivalent to the Calabi-Yau condition, $c_1(X)=0$. Further, we will assume that $X$ is not a direct product manifold, since the structure of Hodge numbers is more complicated in this case. This means that the configuration matrix~\eqref{Q} cannot be brought into a block-diagonal form by any combination of a row and column permutations.\\[2mm]
We would now like to construct the tangent bundle, $TX$ of the manifold $X$ by combining the Euler sequence for the tangent bundle $T{\cal A}$ of the ambient space with the normal bundle sequence. To this end, we need to introduce line bundles on ${\cal A}$ and $X$. For a single projective space $\mathbb{P}^n$, we use the standard notation ${\cal O}_{\mathbb{P}^n}(k)$ for the $k^{\rm th}$ tensor power of the hyperplane bundle. Line bundles on the full ambient space $\cA$ are given by ${\cal O}_{\cal A}({\bf k})={\cal O}_{\mathbb{P}^{n_1}}(k^1)\otimes\cdots\otimes {\cal O}_{\mathbb{P}^{n_m}}(k^m)$, where ${\bf k}=(k^1,\ldots ,k^m)$, and their restrictions to the CICY manifolds are denoted by ${\cal O}_X({\bf k})={\cal O}_{\cal A}({\bf k})|_X$. 

To a configuration matrix~\eqref{Q} we can associate the following sum of line bundles
\beq
\cN~=~ \bigoplus_{a=1}^K \cO_{\cA}({\bf q}_a)
\eeq
whose sections are the defining polynomials ${\bf p}=(p_1,\ldots ,p_K)$. Its restriction $N={\cal N}|_X$ is the normal bundle of $X$ and the associated normal bundle sequence
\beq\label{eq:nbsequence}
0~\longrightarrow~ TX~ \longrightarrow~  T{\cal A}|_X~ \longrightarrow~  N~ \longrightarrow~  0~.
\eeq
gives the tangent bundle $TX$ of $X$ in terms of the normal bundle and the tangent bundle $T{\cal A}$ of the ambient space. This short exact sequence induces a long exact sequence in cohomology which (using that $H^{3,1}(X)=H^{0,1}(X)=0$ for Calabi-Yau three-folds which are not direct products) is explicitly given by
\beq
\begin{array}{lllllllll}
0 & \lrarr &0&\lrarr&H^0(X,T\cA|_X)&\lrarr&H^0(X,N)&\lrarr&\\[3pt]
& \lrarr &H^{2,1}(X)&\lrarr&H^1(X,T\cA|_X)&\lrarr&H^1(X,N)&\lrarr&\\[3pt]
& \lrarr &H^{1,1}(X)&\lrarr&H^2(X,T\cA|_X)&\lrarr&H^2(X,N)&\lrarr&\\[3pt]
& \lrarr &0&\lrarr&H^3(X,T\cA|_X)&\lrarr&H^3(X,N)&\lrarr &0\; .
\end{array}
\eeq
This implies for the cohomology $H^{2,1}(X)$ that
\beq
\begin{aligned}\label{H21seq1}
H^{2,1}(X) \cong \frac{H^0(X,N)}{H^0(X,T\cA|_X)} \oplus {\rm Ker}\left( H^1(X,T\cA|_X)\rightarrow H^1(X,N)\right)~.
\end{aligned}
\eeq
Since $N$ is a sum of line bundles, its cohomology can be relatively easily computed from line bundle cohomology on $\cA$, using the Koszul spectral sequence, as we will discuss below. The tangent bundle $T\cA$ can be obtained from the Euler sequence
\begin{equation}
 0~ \longrightarrow~ {\cal O}_{\cal A}^{\oplus m}\stackrel{f_1}{ \longrightarrow}\, {\mathfrak S}\,\stackrel{f_2}{ \longrightarrow}\, T\cA~ \longrightarrow~ 0\quad\mbox{where}\quad \mathfrak{S}=\bigoplus_{r=1}^m{\cal O}_{\cal A}({\bf e}_r)^{\oplus (n_r+1)}\; ,\label{ESamb}
\end{equation}
where ${\bf e}_r$ are the standard unit vectors in $m$ dimensions. Since the normal bundle sequence~\eqref{eq:nbsequence} actually involves $T{\cal A}|_X$ we require the restriction of the Euler sequence to $X$ which reads
\beq\label{eq:Eulersequence}
0~\longrightarrow ~{\cal O}_X^{\oplus m}\longrightarrow S\longrightarrow T{\cal A}|_X\longrightarrow~ 0\quad\mbox{where}\quad S=\bigoplus_{r=1}^m{\cal O}_{X}({\bf e}_r)^{\oplus (n_r+1)}\; .
\eeq
The associated long exact sequence in cohomology,
\beq
\begin{array}{llcllllll}
0 & \lrarr &H^0(X,{\cal O}_X^{\oplus m})\cong\mathbb{C}^m&\lrarr&H^0(X,S)&\lrarr&H^0(X,T{\cal A}|_X)&\lrarr&\\[3pt]
& \lrarr &0&\lrarr&H^1(X,S)&\lrarr&H^1(X,T{\cal A}|_X)&\lrarr&\\[3pt]
& \lrarr &0&\lrarr&H^2(X,S)&\lrarr&H^2(X,T{\cal A}|_X)&\lrarr&\\[3pt]
& \lrarr &H^3(X,{\cal O}_X^{\oplus m})\cong\mathbb{C}^m&\lrarr&H^3(X,S)&\lrarr&H^3(X,T{\cal A}|_X)&\lrarr &0
\end{array}
\eeq
leads to the identifications
\begin{align}
H^0(X,T\cA|_X) &\cong H^0(X,S))/H^0(X,{\cal O}_X^{\oplus m})\\
H^1(X,T\cA|_X) &\cong H^{1}(X,S)\; ,
\end{align}
where we have used the fact that $H^0(X,{\cal O}_X)\cong H^3(X,{\cal O}_X)\cong \IC$ and that the cohomology groups $H^1(X,{\cal O}_X)$ and $H^2(X,{\cal O}_X)$ are trivial. Combining Eq.~\eqref{H21seq1} with these identification leads to
\beq
\begin{aligned}
H^{2,1}(X) \cong \frac{H^0(X,N)}{ H^0(X,S))/H^0(X,{\cal O}_X^{\oplus m})}\oplus {\rm Ker}\left( H^1(X,S)\rightarrow H^1(X,N)\right)~. \label{h21res}
\end{aligned}
\eeq
It turns out that the kernel in the above equation vanishes for many cases of interest. In particular, this is true for all entries in the standard CICY list~\cite{Candelas:1987kf, Candelas:1987du} with freely-acting symmetries.
Under this assumption, the expression for $H^{2,1}(X)$ simplifies to
\begin{equation}
 H^{2,1}(X) \cong \frac{H^0(X,N)}{ H^0(X,S))/H^0(X,{\cal O}_X^{\oplus m})}\; , \label{h21ressim}
\end{equation} 
and the upstairs Hodge number can be obtained from
\begin{equation} 
 h^{2,1}(X)=h^0(X,N)-h^0(X,S)+m\; . \label{h21up}
\end{equation}
While the upstairs Hodge numbers for CICYs are well-known, a computation along the above lines provides a basic check of our method. The upstairs Euler number, $\eta(X)$, can be computed by elementary methods from the data in the configuration matrix~\eqref{Q}, as explained in Ref.~\cite{hübsch1994calabi}. The other non-trivial Hodge number $h^{1,1}(X)$ can then be obtained from the standard formula
\begin{equation}
 h^{1,1}(X)=\eta(X)/2+h^{2,1}(X)\; .
\end{equation} 

The computation of $H^{2,1}(X)$ outlined above requires the computation of line bundle cohomology on $X$, specifically for the line bundle sums $N$ and $S$. This is accomplished by means of the Koszul sequence which relates cohomology on $X$ to cohomology on the ambient space ${\cal A}$. In general, for any line bundle ${\cal L}$ on ${\cal A}$ and its restriction $L={\cal L}|_X$ the Koszul sequence reads
\begin{equation}
 0~\longrightarrow~ \wedge^K\!\!{\cal N}^*\!\otimes{\cal L}~\stackrel{{\bf p}_{K-1}}{\longrightarrow}~\cdots~\stackrel{{\bf p}_2}{\longrightarrow}~\wedge^2\!\!{\cal N}^*\!\otimes{\cal L}~\stackrel{{\bf p}_1}{\longrightarrow} ~{\cal N}^*\!\otimes{\cal L}~\stackrel{{\bf p}_0={\bf p}}{\longrightarrow}~{\cal L}~\longrightarrow~ L\longrightarrow ~ 0\; ,
\end{equation}
where the map ${\bf p}_0={\bf p}$ is given by the defining polynomials of the CICY $X$ and the higher maps ${\bf p}_1,\ldots,{\bf p}_{K-1}$ are the corresponding induced maps on the anti-symmetric powers of ${\cal N}^*$. Using spectral sequence techniques (see, for example, Refs.~\cite{griffiths2011principles,Distler:1987ee} for an accessible account) this sequence can be used to express the cohomology of $L$ in terms of cohomologies of ambient space line bundle sums $\wedge^k{\cal N}^*\otimes{\cal L}$. Together with standard results for line bundle cohomology on projective spaces~\cite{hartshorne1977algebraic} this allows for an explicit computation of line bundle cohomology on $X$ in terms of ambient space line bundle cohomology. From the two sequences~\eqref{eq:nbsequence} and \eqref{eq:Eulersequence} the relevant line bundle sums on the ambient space are
\begin{equation}
 {\cal O}_{\cal A}^{\oplus m}\;,\qquad {\cal N}= \bigoplus_{a=1}^K \cO_{\cA}({\bf q}_a)\;,\qquad {\mathfrak S}=\bigoplus_{r=1}^m{\cal O}_{\cal A}({\bf e}_r)^{\oplus (n_r+1)}\; .\label{threebundles}
\end{equation} 
All cohomology groups required for the calculation of $H^{2,1}(X)$ from Eq.~\eqref{h21res} can be expressed in terms of ambient space cohomology of the above three line bundle sums and their tensor powers, by means of the Koszul sequence. In particular, we always have~\footnote{For simplicity of notation, here and in the following we omit the first argument, $\cA$, from cohomologies whenever we refer to the ambient space.}
\begin{equation}
 H^0(X,{\cal O}_X^{\oplus m})\cong H^0({\cal O}_\cA)^{\oplus m}\; .
\end{equation}
For the other two required cohomologies, $H^0(X,N)$ and $H^0(X,S)$, the correspondence has to be worked out case by case and we will do this explicitly for the examples in the next section. This concludes the discussion of the upstairs manifold. 

\subsection{The quotient manifold and its Hodge numbers}
Next, we assume that $X$ has a freely-acting discrete symmetry, $G$, of order $|G|$ and we define the quotient manifold $\tilde{X}=X/G$. Our goal is to compute the Hodge numbers $h^{1,1}(\tilde{X})$ and $h^{2,1}(\tilde{X})$ of this quotient. Divisibility of the Euler number means that
\begin{equation}
\eta(\tilde{X})=\eta(X)/|G|\; ,
\end{equation}
so it is sufficient to compute only one of the downstairs Hodge numbers. Starting with the discussion in the previous sub-section, we will set up an algorithm to compute $h^{2,1}(\tilde{X})$. In general, the downstairs cohomology $H^{2,1}(\tilde{X})$ is given by the $G$-invariant part of the upstairs cohomology $H^{2,1}(X)$, so
\begin{equation}
 H^{2,1}(\tilde{X})\cong \left(H^{2,1}(X)\right)_{\rm inv}\; . \label{h21down}
\end{equation} 
Hence, we should work out the equivariant structures on all bundles involved and determine the $G$ representation content of $H^{2,1}(X)$ (for similar work chasing equivariant structures through sequences defining bundles see \cite{Anderson:2009mh}). As we will see, there are three representations of the group $G$ which enter this discussion. The first of these is the (projective/permutation) representation on the homogeneous coordinates of the ambient space, denoted by
\begin{equation}
 \gamma :G\rightarrow S_m\ltimes (PGL(\mathbb{C}^{n_1+1})\times\cdots\times PGL(\mathbb{C}^{n_m+1}))\; . \label{gammadef}
\end{equation} 
Further, we have a representation
\begin{equation}
 \rho : G\rightarrow H^0({\cal A},{\cal N}^*\otimes{\cal N}) \label{rhodef}
\end{equation}
which describes the symmetry action on the defining polynomials or, equivalently, an equivariant structure on the bundle $\cN$. The idea is that the CICY $X$ is invariant under the combined action of $\gamma$ and $\rho$ and these are precisely the representations which are provided by Braun's classification in Ref.~\cite{Braun:2010vc}. The third required representation of $G$ is the permutation representation
\begin{equation}
\pi:G\rightarrow S_m \label{pidef}
\end{equation}
which captures the part of the $\gamma$-action on the homogeneous ambient space coordinates which permutes projective spaces of the same dimension. The representations $\pi$ can be easily obtained from the representations~$\gamma$, as provided in the classification of Ref.~\cite{Braun:2010vc}, by extracting the part of $\gamma$ which permutes entire projective spaces, discarding any non-trivial action on coordinates within each projective space. 

In order to work out the $G$-invariant part of $H^{2,1}(X)$ we require the $G$-representation content of the various cohomologies which appear in the formula~\eqref{h21ressim}. As we have discussed, these cohomologies can, in turn, be expressed in terms of ambient space cohomologies of the three bundles~\eqref{threebundles} and their tensor powers.

The conclusion from this discussion is that the $G$-representation content of all relevant cohomologies is determined once we fix equivariant structures on the three line bundle sums~\eqref{threebundles} which constitute our basic building blocks. Since these three bundles are globally generated an equivariant structure can be specified by a $G$-action on their sections and this turns out to be a convenient way to proceed. The sections can be written as homogeneous polynomials of appropriate degrees in the ambient space coordinates ${\bf x}$ and they are explicitly given by
\begin{equation}
 \Gamma({\cal O}_{\cal A}^{\oplus m})=\{{\bf c}\in\mathbb{C}^m\}\,,\qquad
 \Gamma({\mathfrak S})=\{({\bf l}_1({\bf x}_1),\ldots ,\ldots ,{\bf l}_{m}({\bf x}_m))\}\,,\qquad 
 \Gamma({\cal N})=\bigoplus_{a=1}^K\mathbb{C}[{\bf x}]_{{\bf q}_a}\; ,
\end{equation} 
where ${\bf l}_r=(l_{r,0},\ldots ,l_{r,n_r})$ are $n_r+1$-dimensional vectors of polynomials linear in ${\bf x}_r$ and $\mathbb{C}[{\bf x}]_{\bf k}$ denotes the multi-degree ${\bf k}$ part of the ambient space coordinate ring $\mathbb{C}[{\bf x}]$. A consistent choice of $G$-actions on these sections which leads to the required equivariant structure on ${\cal N}$ and $T{\cal A}$ is given by
\begin{equation}
 R_{ \Gamma({\cal O}_{\cal A}^{\oplus m})}=\pi\,,\qquad
 R_{\Gamma({\mathfrak S})}(g)({\bf l})({\bf x})=\gamma(g){\bf l}\left(\gamma(g)^{-1}{\bf x}\right)\,,\qquad
 R_ {\Gamma({\cal N})}(g)({\bf n})({\bf x})=\rho(g){\bf n}\left(\gamma(g)^{-1}{\bf x}\right)\; . \label{Rreps}
\end{equation} 
We would like to show that this is indeed the correct choice. First, the action of $\gamma(g)^{-1}$ on the argument ${\bf x}$ is the standard way by which $G$ acts on sections. The overall multiplicative action of $G$, on the other hand, corresponds to a choice of equivariant structure and needs to be justified. For the bundle ${\cal N}$ the overall action by  $\rho$ is evidently correct, since $\rho$ provides an equivariant structure on $\cN$.

To discuss the other two bundles we should first introduce the global vector fields
\begin{equation}
 \Gamma(T\cA)=\left\{\left[\sum_{r=1}^m{\bf l}_r({\bf x}_r)\cdot\frac{\partial}{\partial {\bf x}_r}\right]\right\}
\end{equation} 
on $\cA$ where the ${\bf l}_r$ are $(n_r+1)$-dimensional vectors of linear polynomials in ${\bf x}_r$, as before, and the square bracket indicates equivalence classes taken with respect to the subset spanned by ${\bf x}_r\cdot\frac{\partial}{\partial {\bf x}_r}$ for $r=1,\ldots ,m$. We should now look at the Euler sequence~\eqref{ESamb}. The maps $f_1$ and $f_2$ in this sequence induce the following maps
\begin{equation}
 f_1({\bf c})=(c_1{\bf x}_1,\ldots , c_m{\bf x}_m)\qquad\qquad f_2({\bf l})=\left[\sum_{r=1}^m{\bf l}_r({\bf x}_r)\cdot\frac{\partial}{\partial {\bf x}_r}\right]
\end{equation} 
on the sections. Evidently, given the equivalence class taken on the RHS of the second equation, we have $f_2\circ f_1=0$, as should be the case for a complex. We can also verify that
\begin{equation}
 f_1\circ R_{ \Gamma({\cal O}_{\cal A}^{\oplus m})}(g)=R_{\Gamma({\mathfrak S})}(g)\circ f_1\qquad\qquad
 f_2\circ R_{\Gamma({\mathfrak S})}(g)=R_{\Gamma(T\cA)}(g)\circ f_2
\end{equation} 
where $R_{\Gamma(T\cA)}(g)$ is the obvious action of $G$ on the vector fields. This means that the chosen representations intertwine the maps in the Euler sequence for the canonical $G$-action on the vector fields and, therefore, represent the correct choice. Note in particular, that the non-trivial choice of $\pi$ acting on the sections of ${\cal O}_{\cal A}^{\oplus m}$ is required and that the trivial representation $R_{ \Gamma({\cal O}_{\cal A}^{\oplus m})}={\rm id}$ would not satisfy the intertwining conditions.\\[2mm] 
From these equivariant structures and the Koszul sequence, we can work out the $G$-representation content of all relevant cohomologies and obtain the characters
\begin{equation}
 \chi_{H^0(X,{\cal O}_X^{\oplus m})}\,,\qquad \chi_{H^0(X,S)}\,,\qquad \chi_{H^0(X,N)}\; ,
\end{equation} 
of the cohomologies $H^0(X,{\cal O}_X^{\oplus m})$, $H^0(X,S)$ and $H^0(X,N)$. Provided the kernel in Eq.~\eqref{h21res} vanishes the character for $H^{2,1}(X)$ is then given by
\begin{equation}
 \chi_{H^{2,1}(X)}=\chi_{H^0(X,N)}- \chi_{H^0(X,S)}+ \chi_{H^0(X,{\cal O}_X^{\oplus m})}\; . \label{chi21}
\end{equation} 
In general, for a character $\chi$, the number of singlets, $\nu$, can be computed from the formula
\begin{equation} 
 \nu=\frac{1}{|G|}\sum_{g\in G}\chi(g)\; . \label{nuchar}
\end{equation} 
Let us denote by $\nu_{H^0(X,{\cal O}_X^{\oplus m})}$, $\nu_{H^0(X,N)}$, and $\nu_{H^0(X,S)}$ the number of $G$-singlets in the three relevant cohomologies. In practice, these numbers are most easily obtained by demanding invariance under the transformations~\eqref{Rreps} and the corresponding transformations induced on tensor bundles. From Eq.~\eqref{h21down} the downstairs Hodge number $h^{2,1}(\tilde{X})$ equals the number of $G$-singlets in $H^{2,1}(X)$ and hence, by applying Eq.~\eqref{nuchar} to Eq.~\eqref{chi21}, we find that
\begin{equation}
\boxed{
 ~~~h^{2,1}(\tilde{X})=\nu_{H^0(X,N)}-\nu_{H^0(X,S)}+\nu_{H^0(X,{\cal O}_X^{\oplus m})}\; ,~~} \label{h21form}
\end{equation}
provided the kernel in Eq.~\eqref{h21res} vanishes. Eq.~\eqref{h21form} is our key result for the computation of the downstairs Hodge number $h^{2,1}(\tilde{X})$. Given that the index divides, so $\eta(\tilde{X})=\eta(X)/|G|$, the other downstairs Hodge number is easily obtained from
\begin{equation}
 h^{1,1}(\tilde{X})=\frac{\eta(X)}{2|G|}+h^{2,1}(\tilde{X})\; . \label{h11down}
\end{equation} 
\subsection{Summary of algorithm}
We would now briefly like to summarise our algorithm before we discuss a number of explicit applications in the next section.
\begin{itemize}
\item {\bf Set-up}: Define the ambient space and the CICY by providing the configuration matrix~\eqref{Q} and write down the bundles ${\cal O}_{\cal A}^{\oplus m}$, ${\mathfrak S}$ and ${\cal N}$ for this manifold.
\item {\bf Cohomologies}: Compute the cohomologies $H^0(X,{\cal O}_X^{\oplus m})$, $H^0(X,S)$ and $H^0(X,N)$ in terms of cohomologies of the ambient space bundles~\eqref{threebundles} and their tensor powers. (Also check that the kernel in Eq.~\eqref{h21res} vanishes.)
\item {\bf Upstairs Hodge number}: As a basic check, compute the upstairs Hodge number $h^{2,1}(X)$ from Eq.~\eqref{h21up}.
\item {\bf Symmetry}: Define the action of the freely-acting symmetry $G$ by providing the representations $\gamma$ in Eq.~\eqref{gammadef} and $\rho$ in Eq.~\eqref{rhodef}. Also compute the permutation representation $\pi$ in Eq.~\eqref{pidef}.
\item {\bf Singlets}: Compute, in turn, the number of $G$-singlets in $H^0(X,{\cal O}_X^{\oplus m})$, $H^0(X,S)$ and $H^0(X,N)$.
\item {\bf Downstairs Hodge numbers}: Compute $h^{2,1}(\tilde{X})$ from Eq.~\eqref{h21form} and $h^{1,1}(\tilde{X})$ from Eq.~\eqref{h11down}.
\end{itemize}

\section{Explicit examples}
In this section, we will explicitly illustrate the above algorithm by computing the downstairs Hodge numbers for a number of CICYs. The CICY data is taken from the standard list~\cite{Candelas:1987kf, Candelas:1987du} and the freely acting symmetries are taken from Braun's classification in Ref.~\cite{Braun:2010vc}. The relevant CICY data required for this paper is available at the website~\cite{CicySymm}. This includes the configuration matrices for the relevant CICYs, their identifying number which gives their position in the original list of Ref.~\cite{Candelas:1987kf, Candelas:1987du}, the upstairs Hodge numbers $h^{1,1}(X)$, $h^{2,1}(X)$ and a list of symmetries, each specified by the matrices $\gamma(g)$, $\rho(g)$ (see Eqs.~\eqref{gammadef}, \eqref{rhodef}). If a CICY has more than one symmetry we will refer to a specific symmetry by its position in this list.

\vspace{21pt}
\subsection{Example 1: A $\mathbb{Z}_4$ symmetry on the tetra-quadric}
\underline{\bf Set-up}: We consider the tetra-quadric CICY with number 7862, defined as the zero locus of a multi-degree ${\bf q}=(2,2,2,2)$ polynomial $p$ in the ambient space ${\cal A}=\mathbb{P}^1\times\mathbb{P}^1\times\mathbb{P}^1\times\mathbb{P}^1$. The configuration matrix is given by
\begin{equation}
 X=\left[\begin{array}{c|c}\mathbb{P}^1&2\\\mathbb{P}^1&2\\\mathbb{P}^1&2\\\mathbb{P}^1&2\end{array}\right]^{4,68}_{-128}\; ,
\end{equation} 
and the three relevant ambient space line bundle sums are
\begin{equation}
 {\cal O}_{\cal A}^{\oplus 4}\,,\qquad {\mathfrak S}={\cal O}_{\cal A}({\bf e}_1)^{\oplus 2}\oplus\cdots {\cal O}_{\cal A}({\bf e}_4)^{\oplus 2}\,,\qquad
 {\cal N}={\cal O}_{\cal A}(2,2,2,2)\; .
\end{equation}\\[-2mm] 
\underline{\bf Cohomologies}: For the bundles $S$ and $N$, the long exact sequences associated to their Koszul sequences are given by
\begin{equation}
\begin{array}{l}\\h^0(\cdot)\\h^1(\cdot)\\h^2(\cdot)\\h^3(\cdot)\end{array}
\qquad
 \begin{array}{ccccc}
  {\cal N}^*\otimes{\mathfrak S}&\stackrel{p}{\rightarrow}&{\mathfrak S}&\rightarrow&S\\
   0&&16&&16\\
   0&&0&&0\\
   0&&0&&0\\
   0&&0&&0
\end{array}
\qquad\qquad\qquad
 \begin{array}{ccccc}
  {\cal O}_{\cal A}&\stackrel{p}{\rightarrow}&{\cal N}&\rightarrow&N\\
   1&&81&&80\\
   0&&0&&0\\
   0&&0&&0\\
   0&&0&&0
\end{array}
\end{equation}
where we have omitted the zeros to the left and right of these sequences. This shows that
\begin{equation}
  H^0(X,S)\cong H^0(\mathfrak{S})\;,\qquad H^0(X,N)\cong \frac{H^0({\cal N})}{p(H^0({\cal O}_{\cal A}))}\; .
\end{equation}  
Further, since all higher cohomologies of $N$ and $S$ are zero, the kernel in Eq.~\eqref{h21res} vanishes.\\[7mm]
\underline{\bf Upstairs Hodge number}: From the previous results we conclude that $h^0(X,S)=16$ and $h^0(X,N)=80$ which implies for the upstairs Hodge number
\begin{equation}
 h^{2,1}(X)=h^0(X,N)-h^0(X,S)+4=80-16+4=68\; ,
\end{equation}
in line with expectations.\\[7mm] 
\underline{\bf Symmetry}: We would like to consider the second freely-acting symmetry of the tetra-quadric which corresponds to the group $G=\mathbb{Z}_4$ with generator $g$. It is defined by the representations
\begin{equation}
\gamma (g)=\left(\begin{array}{llll}0&\sigma_3&0&0\\{\bf 1}_2&0&0&0\\0&0&0&\sigma_3\\0&0&{\bf 1}_2&0\end{array}\right)\,,\qquad \rho(g)=1\,,\qquad
\sigma_3={\rm diag}(1,-1)\; . \label{ex1gamma}
\end{equation}
The associated permutation representation $\pi$ acts on the four $\mathbb{P}^1$ factors as
\begin{equation}
 \pi(g)=\left(\begin{array}{llll}0&1&0&0\\1&0&0&0\\0&0&0&1\\0&0&1&0\end{array}\right)\; . \label{piex1}
\end{equation}\\[2mm] 
\underline{\bf Singlets}: {\bf a)} First, we compute the number of singlets in $H^0(X,{\cal O}_X^{\oplus m})$ by solving the constraint $\pi(g){\bf c}={\bf c}$ for a four-dimensional complex vector ${\bf c}=(c_1,c_2,c_3,c_4)^T$ and $\pi(g)$ as given in Eq.~\eqref{piex1}. Clearly, the solution space is two-dimensional so that
\begin{equation} 
 \nu_{H^0(X,{\cal O}_X^{\oplus m})}=2\; .
\end{equation}\\[-2mm]
{\bf b)} In order to compute the number of singlets in $H^0(X,S)\cong H^0({\cal A},{\mathfrak S})$ we first observe that the matrix~\eqref{ex1gamma} splits into two $4\times 4$ blocks with the same structure. It is, therefore, sufficient to calculate for one of these blocks. From Eq.~\eqref{Rreps}, the action of the representation $R_{\Gamma({\mathfrak S})}$ (for one of the blocks) is
\begin{equation}
 R_{\Gamma({\mathfrak S})}(g)\left(\begin{array}{c}c_0x_0+c_1x_1\\\tilde{c}_0x_0+\tilde{c}_1x_1\\d_0y_0+d_1y_1\\\tilde{d}_0y_0+\tilde{d}_1y_1\end{array}\right)=
 \left(\begin{array}{c}d_0x_0-d_1x_1\\-\tilde{d}_0x_0+\tilde{d}_1x_1\\c_0y_0+c_1y_1\\\tilde{c}_0y_0+\tilde{c}_1y_1\end{array}\right)\; .
\end{equation} 
For a singlet, the RHS needs to equal the argument on the LHS which implies $c_0=d_0$, $\tilde{c}_1=\tilde{d}_1$ and the vanishing of all other coefficients. This means we have two singlets in each $4\times 4$ block, for a total of
\begin{equation}
 \nu_{H^0(X,S)}=4\; .
\end{equation}\\[-2mm] 
{\bf c)} Finally, we require the number of $G$-singlets in $H^0(X,N)$ which is simply given by the number of $G$-invariant tetra-quadric minus one (corresponding to the defining polynomial which has to be taken off due to the quotient in $H^0(X,N)\cong H^0({\cal A},{\cal N})/p(H^0({\cal A},{\cal O}_{\cal A}))$). The number of $G$-invariant tetra-quadrics is $21$, hence,
\begin{equation}
 \nu_{H^0(X,N)}=20\; .
\end{equation}\\[2mm]
\underline{\bf Downstairs Hodge numbers}: Altogether, from Eq.~\eqref{h21form}, this lead to the downstairs Hodge number
\begin{equation}
 h^{2,1}(\tilde{X})= \nu_{H^0(X,N)}- \nu_{H^0(X,S)}+ \nu_{H^0(X,{\cal O}_X^{\oplus m})}=20-4+2=18\; .
\end{equation}
From Eq.~\eqref{h11down}, the other downstairs Hodge number is given by
\begin{equation}
 h^{1,1}(\tilde{X})=\eta(X)/8+h^{2,1}(\tilde{X})=-128/8+18=2\; .
\end{equation} 
These results agree with the ones obtained in Refs.~\cite{Candelas:2008wb, Candelas:2015amz}.  

\vspace{21pt}
\subsection{Example 2: A co-dimension two CICY with a $\mathbb{Z}_2$ symmetry}
\underline{\bf Set-up}: This co-dimension two CICY carries the number 2565 and has only a single freely-acting symmetry with group $G=\mathbb{Z}_2$. The Hodge numbers for this quotient have not been computed before. The manifold is defined in the ambient space ${\cal A}=\mathbb{P}^1\times\mathbb{P}^1\times\mathbb{P}^1\times \mathbb{P}^2$ with configuration matrix
\begin{equation}
 X=\left[\begin{array}{l|ll}\mathbb{P}^1&0&2\\\mathbb{P}^1&2&0\\\mathbb{P}_1&2&0\\\mathbb{P}^2&1&2\end{array}\right]^{10,26}_{-32}\; .
\end{equation} 
The relevant ambient space bundles are ${\cal O}_{\cal A}^{\oplus 4}$ and
\begin{eqnarray}
 {\mathfrak S}&=&{\cal O}_{\cal A}({\bf e}_1)^{\oplus 2}\oplus{\cal O}_{\cal A}({\bf e}_2)^{\oplus 2}\oplus{\cal O}_{\cal A}({\bf e}_3)^{\oplus 2}\oplus{\cal O}_{\cal A}({\bf e}_1)^{\oplus 3}\\
 {\cal N}&=&{\cal N}_1\oplus{\cal N}_2={\cal O}_{\cal A}(0,2,2,1)\oplus{\cal O}_{\cal A}(2,0,0,2)\; .
\end{eqnarray}\\[2mm] 
\underline{\bf Cohomologies}: The Koszul sequence for $S={\mathfrak S}|_X$ shows that
\begin{equation}
 H^0(X,S)\cong H^0({\mathfrak S})=\underbrace{H^0({\cal A}_{\cal A}({\bf e}_1))^{\oplus 2}}_{\text{ 4 dim.}}\oplus \underbrace{H^0({\cal A}_{\cal A}({\bf e}_2))^{\oplus 2}}_{\text{ 4 dim.}}\oplus \underbrace{H^0({\cal A}_{\cal A}({\bf e}_3))^{\oplus 2}}_{\text{ 4 dim.}}\oplus \underbrace{H^0({\cal A}_{\cal A}({\bf e}_4))^{\oplus 3}}_{\text{9 dim.}}\; ,\label{ex4H0S}
\end{equation}
so that
\begin{equation}
 h^0(X,S)=21\; .
\end{equation} 
For the normal bundle $N={\cal N}|_X$ the Koszul sequence leads to
\begin{equation}
 H^0(X,N)\cong\frac{H^0({\cal N})}{H^0({\cal N}^*\otimes {\cal N})}\; ,
\end{equation} 
where
\begin{eqnarray}
 H^0({\cal N})&=&\underbrace{H^0({\cal N}_1)}_{\text{27 dim.}}\oplus \underbrace{H^0({\cal N}_2)}_{\text{18 dim.}}\\[8pt]
 H^0({\cal N}^*\otimes{\cal N})&=&\underbrace{H^0({\cal N}_1^*\otimes {\cal N}_1)}_{\text{1 dim.}}\oplus \underbrace{H^0({\cal N}_2^*\otimes {\cal N}_2)}_{\text{1 dim.}}=H^0({\cal O}_{\cal A})^{\oplus 2}\; .
\end{eqnarray} 
Hence, we have
\begin{equation}
 h^0(X,N)=h^0({\cal N})-h^0({\cal N}^*\otimes{\cal N})=45-2=43\; .
 \end{equation}
Further, it turns out that $h^1(X,S)=3$ and $h^1(X,N)=9$. Even though both of these cohomologies are non-trivial it can be checked that the map between them is injective and, hence, that the kernel in Eq.~\eqref{h21res} vanishes.\\[2mm] 
\underline{\bf Upstairs Hodge number}:  Combining the above results, we find for the upstairs Hodge number 
\begin{equation}
 h^{2,1}(X)=h^0(X,N)-h^0(X,S)+4=43-21+4=26\; ,
\end{equation}
in line with expectations.\\[2mm]
\underline{\bf Symmetry}: The relevant representations of the symmetry group $G=\mathbb{Z}_2$ with generator $g$ are
\begin{eqnarray}
 \gamma(g)&=&{\rm diag}(-1,1,-1,1,-1,1,-1,-1,1)\\
 \rho(g)&=&{\rm diag}(-1,1)\\
 \pi(g)&=&\mathbf{1}_4\; .
\end{eqnarray}\\[2mm] 
\underline{\bf Singlets}: {\bf a)} Since the $\pi$-action is trivial it is immediately clear that
\begin{equation}
 \nu_{H^0(X,{\cal O}_X^{\oplus 4})}=4\; .
\end{equation}\\[2mm]
{\bf b)} We denote the projective ambient space coordinates by $((x_0,x_1),(y_0,y_1),(z_0,z_1),(t_0,t_1,t_2))$. From Eq.~\eqref{ex4H0S}, the cohomology $H^0(X,S)$  can be represented by a vector
\begin{eqnarray}
 {\bf l}({\bf x})&=&(a_1x_0+b_1x_1,a_1'x_0+b_1'x_1,a_2y_0+b_2y_1,a_2'y_0+b_2'y_1,a_3z_0+b_3z_1,a_3'z_0+b_3'z_1,\\
 &&\;\,a_4t_0+b_4t_1+c_4t_2,a_4't_0+b_4't_1+c_4't_2,a_4''t_0+b_4''t_1+c_4''t_2)^T\; .
\end{eqnarray}
with a total of $h^0(X,S)=21$ arbitrary coefficients. Applying to this vector the constraint ${\bf l}({\bf x})=\gamma(g){\bf l}(\gamma(g)^{-1}{\bf x})$ we learn that the number of invariants is
\begin{equation}
 \nu_{H^0(X,S)}=11\; .
\end{equation}\\[2mm]
{\bf c)} For $H^0(X,N)$ we require the representations 
\begin{eqnarray}
 H^0({\cal N}_1)&=&H^0({\cal O}_{\cal A}(0,2,2,1))={\rm Span}(y_0^2,y_0y_1,y_1^2)\otimes{\rm Span}(z_0^2,z_0z_1,z_1^2)\otimes {\rm Span}(t_0,t_1,t_1)\\
 H^0({\cal N}_2)&=&H^0({\cal O}_{\cal A}(2,0,0,2))={\rm Span}(x_0^2,x_0x_1,x_1^2)\otimes {\rm Span}(t_0^2,t_0t_1,t_1^2)\\
 H^0({\cal N}^*\otimes {\cal N})&=&H^0({\cal O}_{\cal A})^{\oplus 2}
\end{eqnarray} 
Finding the invariants by a straightforward application of the last Eq.~\eqref{Rreps} leads to
\begin{equation}
 \nu_{H^0({\cal N}_1)}=14\;,\qquad  \nu_{H^0({\cal N}_2)}=10\; ,\qquad \nu_{H^0({\cal N}^*\otimes{\cal N})}=2\; ,
\end{equation} 
and, hence,
\begin{equation}
 \nu_{H^0(X,N)}= \nu_{H^0({\cal N}_1)}+ \nu_{H^0({\cal N}_2)}-\nu_{H^0({\cal N}^*\otimes{\cal N})}=14+10-2=22\; .
\end{equation}\\[2mm] 
\underline{\bf Downstairs Hodge numbers}:  Altogether, this leads to the downstairs Hodge numbers
\begin{eqnarray}
 h^{2,1}(\tilde{X})&=&\nu_{H^0(X,N)}-\nu_{H^0(X,S)}+\nu_{H^0(X,{\cal O}_X^{\oplus 4})}=22-11+4=15\\
 h^{1,1}(\tilde{X})&=&\eta(X)/4+ h^{2,1}(\tilde{X})=-32/4+15=7\; .
\end{eqnarray} 

\vspace{21pt}
\subsection{Example 3: A co-dimension three CICY with $\mathbb{Z}_2\times \mathbb{Z}_2$ symmetry}
\underline{\bf Set-up}: The CICY with number 2568 is a co-dimension three manifold in the ambient space ${\cal A}=(\mathbb{P}^1)^{\times 6}$, defined by the configuration matrix
\begin{equation}\label{eq:dP4_2568}
X=
\cicy{\IP^1 \\ \IP^1\\ \IP^1\\ \IP^1\\ \IP^1\\ \IP^1}
{ ~1 &1 & 0  ~\\
  ~1 &1 & 0  ~\\
  ~0 &2 & 0 ~\\
  ~1 &0 & 1 ~\\
  ~1 &0 & 1 ~\\
    ~0 &0 & 2~\\}_{-32}^{12,28}\; .
\end{equation}
Reading off from the columns of this matrix, the bundle $\cN$ is explicitly given by
\begin{equation}
 {\cal N}={\cal N}_1\oplus{\cal N}_2\oplus{\cal N}_3=\cO_\cA(1,1,0,1,1,0)\oplus\cO_\cA(1,1,2,0,0,0)\oplus\cO_\cA(0,0,0,1,1,2)\; .
\end{equation} 
The other two relevant ambient space bundles are
\begin{equation}
 {\cal O}_{\cal A}^{\oplus 6}\,,\qquad {\mathfrak S}={\cal O}_{\cal A}({\bf e}_1)^{\oplus 2}\oplus\cdots\oplus{\cal O}_{\cal A}({\bf e}_6)^{\oplus 2}\; .
\end{equation} \\[2mm]
\underline{\bf Cohomologies}:  The Koszul sequence for $N$ can be broken up into the three short exact sequences 
\begin{equation}
\arraycolsep=1pt\def\arraystretch{0.7}
\begin{array}{l}\\[0pt]h^0(\cdot )\\[3pt]h^1(\cdot )\\[3pt]h^2(\cdot )\\[3pt]h^3(\cdot )\\[3pt]h^4(\cdot )\\[3pt]h^5(\cdot)\\[3pt]h^6(\cdot )\end{array}\qquad
 \begin{array}{ccccc}
  \wedge^3{\cal N}^*\otimes{\cal N}&\rightarrow&\wedge^2{\cal N}^*\otimes{\cal N}&\rightarrow&K_2\\[4pt]
   0&&0&&0\\[4pt]
   0&&0&&0\\[4pt]
   0&&1&&1\\[4pt]
   0&&6&&6\\[4pt]
   0&&0&&0\\[4pt]
   0&&0&&0\\[4pt]
   0&&0&&0\\[4pt]
\end{array}\qquad\qquad
 \begin{array}{ccccc}
  K_2&\rightarrow&{\cal N}^*\otimes{\cal N}&\rightarrow&K_1\\[4pt]
   0&&3&&3\\[4pt]
   0&&8&&9\\[4pt]
   1&&0&&6\\[4pt]
   6&&0&&0\\[4pt]
   0&&0&&0\\[4pt]
   0&&0&&0\\[4pt]
   0&&0&&0\\[4pt]
\end{array}\qquad\qquad
 \begin{array}{ccccc}
  K_1&\rightarrow&{\cal N}&\rightarrow&N\\[4pt]
   3&&40&&46\\[4pt]
   9&&0&&6\\[4pt]
   6&&0&&0\\[4pt]
   0&&0&&0\\[4pt]
   0&&0&&0\\[4pt]
   0&&0&&0\\[4pt]
   0&&0&&0\\[4pt]
\end{array}
\end{equation}
where $K_1$ and $K_2$ are suitable co-kernels. Combining the information from these sequences we learn that
\begin{eqnarray}
 H^0(X,N)&\cong&\frac{H^0({\cal N})}{H^0({\cal N}^*\otimes {\cal N})}\oplus H^1({\cal N}^*\otimes {\cal N})\oplus H^2(\wedge^2{\cal N}^*\otimes {\cal N})\nonumber\\
 &=&\frac{H^0({\cal N})}{H^0({\cal O}_{\cal A}^{\oplus 3})}\oplus H^1({\cal N}_2\otimes{\cal N}_1^*\oplus {\cal N}_2\otimes{\cal N}_3^*)\oplus H^2({\cal N}_2\otimes {\cal N}_1^*\otimes{\cal N}_3^*)\nonumber\\
 &=&\underbrace{\frac{H^0({\cal N})}{H^0({\cal O}_{\cal A}^{\oplus 3})}}_{\text{40-3=37 dim.}}{\oplus} \underbrace{H^1({\cal O}_{\cal A}{(}{-}{2}{,}{0}{,}{0}{,}{0}{,}{1}{,}{1}{)}{\oplus} {\cal O}_{\cal A}{(}{0}{,}{-}{2}{,}{1}{,}{1}{,}{0}{,}{0}{)})}_{\text{4+4=8 dim.}}{\oplus} \underbrace{H^2({\cal O}_{\cal {A}}{(}{-}{2}{,}{-}{2}{,}{0}{,}{0}{,}{0}{,}{0}{)})}_{\text{1 dim.}}\label{H0Nex2}\\
 h^0(X,N)&=&(40-3)+(4+4)+1=46
\end{eqnarray} 
The situation is much simpler for $S$ whose only non-zero cohomology is
\begin{equation}
 H^0(X,S)\cong \underbrace{H^0({\mathfrak S})}_{\text{24 dim.}}\; . \label{SH0ex3}
\end{equation}
Since $H^1(X,S)=0$ the kernel in Eq.~\eqref{h21res} vanishes.\\[4mm]
\underline{\bf Upstairs Hodge number}:  From the above cohomologies, we have
\begin{equation}
 h^{2,1}(X)=h^0(X,N)-h^0(X,S)+6=46-24+6=28
\end{equation}
which is the correct result.\\[4mm] 
\underline{\bf Symmetry}: We consider the symmetry with number 25 which corresponds to the group $G=\mathbb{Z}_2\times \mathbb{Z}_2$. The relevant representation matrices of the generators $g_1$ and $g_2$ are
\begin{equation}
\begin{array}{llll}
&\gamma(g_1)={\rm diag}(\sigma_3,\sigma_3,\sigma_3,-\sigma_3,\sigma_3,\sigma_3)&\qquad&
\gamma(g_2)={\rm diag}({\bf 1}_2\times \sigma_1,\sigma_1\times {\bf 1}_2,{\bf 1}_2\times\sigma_1)\\
&\rho(g_1)={\bf 1}_3&\qquad& \rho(g_2)={\rm diag}(1,-1,1)\\
&\pi(g_1)={\bf 1}_6&\qquad& \pi(g_2)={\rm diag}({\bf 1}_2,\sigma_1,{\bf 1}_2)\; .
\end{array}
\end{equation}\\[2mm]
\underline{\bf Singlets}: {\bf a)} In order to compute the number of singlets in $H^0(X,{\cal O}_X^{\oplus 6})$ we impose the constraints $\pi(g_1){\bf c}={\bf c}$ and $\pi(g_2){\bf c}={\bf c}$ on an arbitrary six-dimensional complex vector ${\bf c}$ which, obviously, leads to a five-dimensional space. Hence
\begin{equation}
 \nu_{H^0(X,{\cal O}_X^{\oplus 6})}=5\; .
\end{equation}\\[2mm]
{\bf b)} Next, we need to find the $G$-singlets in $H^0(X,S)\cong H^0({\cal A},{\mathfrak S})$. We can split up this problem by first considering the first and last two $\mathbb{P}^1$ factors which are not permuted under $\pi$. Each of these four factors is similar so we can focus on the first with coordinates $u_0,u_1$. Using the action defined in Eq.~\eqref{Rreps}, invariance under $g_1$ in the first factor implies the constraint
\begin{equation}
 \left(\begin{array}{c}au_0+bu_1\\\tilde{a}u_0+\tilde{b}u_1\end{array}\right)= \left(\begin{array}{c}au_0-bu_1\\-\tilde{a}u_0+\tilde{b}u_1\end{array}\right)
\end{equation}
on the two linears involved. It follows that $b=0$ and $\tilde{a}=0$. Acting with the constraint from $g_2$ on the remaining degrees of freedom gives
\begin{equation}
 \left(\begin{array}{c}au_0\\\tilde{b}u_1\end{array}\right)= \left(\begin{array}{c}\tilde{b}u_0\\au_1\end{array}\right)\; ,
\end{equation}
which leads to $a=\tilde{b}$ and, hence, one invariant. This makes for a total of four invariants from the four $\mathbb{P}^1$ factors invariant under $\pi$. For the remaining third and fourth $\mathbb{P}^1$ with coordinates $w_0,w_1,x_0x_1$ the invariant constraint from $g_1$ reads
\begin{equation}
 \left(\begin{array}{c}aw_0+bw_1\\\tilde{a}w_0+\tilde{b}w_1\\cx_0+dx_1\\\tilde{c}x_0+\tilde{d}x_1\end{array}\right)=
  \left(\begin{array}{c}aw_0-bw_1\\-\tilde{a}w_0+\tilde{b}w_1\\cx_0-dx_1\\-\tilde{c}x_0+\tilde{d}x_1\end{array}\right)
\end{equation}
leading to $b=\tilde{a}=d=\tilde{c}=0$. For the remaining vector, the constraint from $g_2$ takes the form
\begin{equation}
 \left(\begin{array}{c}aw_0\\\tilde{b}w_1\\cx_0\\\tilde{d}x_1\end{array}\right)=
 \left(\begin{array}{c}cw_0\\\tilde{d}w_1\\ax_0\\\tilde{b}x_1\end{array}\right)\; ,
\end{equation}
which implies $a=c$ and $\tilde{b}=\tilde{d}$, leaving us with two invariants. Altogether, this means
\begin{equation} 
 \nu_{H^0(X,S)}=4\times 1+2=6\; .
\end{equation}\\[-2mm]  
{\bf c)} Finally, we need to count the number of singlets in $H^0(X,N)$, using the decomposition~\eqref{H0Nex2}. We know that there are $11$ polynomial invariants so that $\nu_{H^0({\cal A},{\cal N})}=11$. Clearly, we have $\nu_{H^0({\cal A},{\cal O}_{\cal A}^{\oplus 3})}=3$. Let us denote the homogeneous coordinates for the six $\mathbb{P}^1$ factors by $((u_0,u_1),(v_0,v_1),(w_0,w_1),(x_0,x_1),(y_0,y_1),(z_0,z_1))$. Starting with $H^1({\cal A},{\cal O}_{\cal A}{(}{-}{2}{,}{0}{,}{0}{,}{0}{,}{1}{,}{1}{)})$ we can represent this cohomology as
\begin{equation}
 H^1({\cal A},{\cal O}_{\cal A}{(}{-}{2}{,}{0}{,}{0}{,}{0}{,}{1}{,}{1}{)})\cong\frac{1}{u_0u_1}{\rm Span}(y_0z_0,y_0z_1,y_1z_0,y_1z_1)\; .
\end{equation} 
Within this four-dimensional space, there is precisely one $G$-singlet given by
\begin{equation}
  \frac{1}{u_0u_1}(y_0z_1+y_1z_0)
\end{equation}
Here, we have taken into account the anti-symmetric nature of the pre-factor $1/(u_0u_1)$ which makes it odd under both $g_1$ and $g_2$ and the negative sign for $g_2$ from the equivariant structure $\rho$ in Eq.~\eqref{rhodef}, given that the line bundle in question is a tensor product which contains ${\cal N}_2$.  This means that
$\nu_{H^1({\cal A},{\cal O}_{\cal A}{(}{-}{2}{,}{0}{,}{0}{,}{0}{,}{1}{,}{1}{)})}=1$. The next contribution, $H^1({\cal A},{\cal O}_{\cal A}{(}{0}{,}{-}{2}{,}{1}{,}{1}{,}{0}{,}{0}{)})$, can be represented as
\begin{equation}
 H^1({\cal A},{\cal O}_{\cal A}{(}{0}{,}{-}{2}{,}{1}{,}{1}{,}{0}{,}{0}{)})\cong \frac{1}{v_0v_1}{\rm Span}(w_0x_0,w_0x_1,w_1x_0,w_1x_1)\; .
\end{equation}
This contains two $G$-invariants, namely
\begin{equation}
 \frac{1}{v_0v_1}w_0x_0\;,\qquad  \frac{1}{v_0v_1}w_1x_1\; ,
\end{equation}
so that $\nu_{H^1({\cal A},{\cal O}_{\cal A}{(}{0}{,}{-}{2}{,}{1}{,}{1}{,}{0}{,}{0}{)})}=2$. The final contribution in Eq.~\eqref{H0Nex2} to be considered is
\begin{equation}
 H^2({\cal A},{\cal O}_{\cal {A}}{(}{-}{2}{,}{-}{2}{,}{0}{,}{0}{,}{0}{,}{0}{)})\cong\frac{1}{u_0u_1v_0v_1}\mathbb{C}\; .
\end{equation} 
This representative is not $G$-invariant and, hence, there is no contribution from this part. Altogether, we have
\begin{equation}
 \nu_{H^0(X,N)}=\nu_{H^0({\cal A},{\cal N})}-\nu_{H^0({\cal A},{\cal N}^*\otimes{\cal N})}+\nu_{H^1({\cal A},{\cal N}^*\otimes{\cal N})}+\nu_{H^2({\cal A},\wedge^2{\cal N}^*\otimes{\cal N})}=11-3+3+0=11\; .
\end{equation} \\[-2mm]
\underline{\bf Downstairs Hodge numbers}:  Combining the above results, we finally find for the downstairs Hodge numbers
\begin{eqnarray}
h^{2,1}(\tilde{X})&=&\nu_{H^0(X,N)}-\nu_{H^0(X,S)}+\nu_{H^0(X,{\cal O}_X^{\oplus m})}=11-6+5=10\\
h^{1,1}(\tilde{X})&=&\eta(X)/8+h^{2,1}(\tilde{X})=-32/8+10=6
\end{eqnarray}
which agrees with the results in Ref.~\cite{Candelas:2015amz}.

\vspace{21pt}
\subsection{Example 4: A co-dimension three CICY with a $\mathbb{Z}_4$ symmetry}
\underline{\bf Set-up}: This example is for the CICY with number 2568, the same as in Example 3. The basic set-up and the computation of cohomologies is identical to Example 3.\\[4mm]
\underline{\bf Symmetry}: We consider the $8^{\rm th}$ symmetry of this manifold, a $G=\mathbb{Z}_4$ symmetry with generator $g$ and associated representations
\begin{eqnarray}
 \gamma(g)&=&\scriptsize\left(
\begin{array}{cccccccccccc}
 0 & 0 & 1 & 0 & 0 & 0 & 0 & 0 & 0 & 0 & 0 & 0 \\
 0 & 0 & 0 & -1 & 0 & 0 & 0 & 0 & 0 & 0 & 0 & 0 \\
 1 & 0 & 0 & 0 & 0 & 0 & 0 & 0 & 0 & 0 & 0 & 0 \\
 0 & 1 & 0 & 0 & 0 & 0 & 0 & 0 & 0 & 0 & 0 & 0 \\
 0 & 0 & 0 & 0 & 0 & 0 & 0 & 0 & 1 & 0 & 0 & 0 \\
 0 & 0 & 0 & 0 & 0 & 0 & 0 & 0 & 0 & 1 & 0 & 0 \\
 0 & 0 & 0 & 0 & 0 & 0 & 0 & 0 & 0 & 0 & 1 & 0 \\
 0 & 0 & 0 & 0 & 0 & 0 & 0 & 0 & 0 & 0 & 0 & 1 \\
 0 & 0 & 0 & 0 & 0 & 0 & 1 & 0 & 0 & 0 & 0 & 0 \\
 0 & 0 & 0 & 0 & 0 & 0 & 0 & 1 & 0 & 0 & 0 & 0 \\
 0 & 0 & 0 & 0 & 1 & 0 & 0 & 0 & 0 & 0 & 0 & 0 \\
 0 & 0 & 0 & 0 & 0 & 1 & 0 & 0 & 0 & 0 & 0 & 0 \\
\end{array}
\right)\\[8pt]
\rho(g)&=&\left(
\begin{array}{ccc}
 0 & 0 & 1 \\
 0 & 1 & 0 \\
 1 & 0 & 0 \\
\end{array}
\right)\qquad
\pi(g)=\left(
\begin{array}{cccccc}
 0 & 1 & 0 & 0 & 0 & 0 \\
 1 & 0 & 0 & 0 & 0 & 0 \\
 0 & 0 & 0 & 0 & 1 & 0 \\
 0 & 0 & 0 & 0 & 0 & 1 \\
 0 & 0 & 0 & 1 & 0 & 0 \\
 0 & 0 & 1 & 0 & 0 & 0 \\
\end{array}
\right)\; .
\end{eqnarray}
\underline{\bf Singlets: a)} We start with $H^0(X,{\cal O}_X^{\oplus 6})$ which we represent by a six-dimensional complex vector ${\bf c}$. Imposing $\pi(g){\bf c}={\bf c}$ shows that
\begin{equation}
 \nu_{H^0(X,{\cal O}_X^{\oplus 6})}=2\; .
\end{equation}\\[-2mm]
{\bf b)} From Eq.~\eqref{SH0ex3}, the cohomology $H^0(X,S)$ can be represented by a $12$-dimensional vector ${\bf l}$ of linears in the appropriate coordinates and imposing ${\bf l}({\bf x})=\gamma(g){\bf l}(\gamma(g)^{-1}{\bf x})$ shows that the number of singlets is given by
\begin{equation}
 \nu_{H^0(X,S)}=6\; .
\end{equation}\\[-2mm] 
{\bf c)} For $H^0(X,N)$ we need to go through the various pieces which appear in Eq.~\eqref{H0Nex2} beginning with $H^0({\cal N})$. Representing this by a three-dimensional vector ${\bf v}$, containing polynomials of degrees as given by the configuration matrix and imposing ${\bf v}({\bf x})=\rho(g){\bf v}(\gamma(g)^{-1}{\bf x})$ gives
\begin{equation}
 \nu_{H^0({\cal N})}=13\; .
\end{equation}
 Elements of $H^0({\cal N}^*\otimes{\cal N})$ can be represented by diagonal matrices $M={\rm diag}(a,b,c)$ and imposing $\rho(g)^\dagger M\rho(g)=M$ shows that
\begin{equation}
 \nu_{H^0({\cal N}^*\otimes{\cal N})}=2\; .
\end{equation}
Denoting the projective ambient space coordinates by ${\bf x}=((x_0,x_1),(y_0,y_1),(z_0,z_1),(t_0,t_1),(u_0,u_1),(v_0,v_1))$, the cohomology $H^1({\cal N}^*\otimes{\cal N})$ can be represented by two-dimensional vectors of the form
\begin{equation}
 {\bf v}({\bf x})=\left(\begin{array}{l}\frac{1}{x_0x_1}(a_1u_0v_0+a_2u_0v_1+a_3u_1v_0+a_4u_1v_1)\\\frac{1}{y_0y_1}(b_1z_0t_0+b_2z_0t_1+b_3z_1t_0+b_4z_1t_1)\end{array}\right)\; .
\end{equation}
 Under the induced action of $\rho$ the two entries of this vector are exchanged (since $\rho$ exchanges ${\cal N}_1$ with ${\cal N}_3$ while leaving ${\cal N}_2$ invariant). Performing this transformation, together with the action of $\gamma(g)^{-1}$ on the coordinates and demanding invariance as usual leads to 
\vspace{-12pt} 
\begin{equation}
 \nu_{H^1({\cal N}^*\otimes{\cal N})}=1\; .
\vspace{-8pt} 
\end{equation}
The final piece is 
\begin{equation}
 H^2(\wedge^2{\cal N}^*\otimes{\cal N})\cong{\rm Span}\left(\frac{1}{x_0x_1y_0y_1}\right)\; .
\end{equation}
This is odd under the action of $g$ and, hence,
$ \nu_{H^2(\wedge^2{\cal N}^*\otimes{\cal N})}=0 $.
Combining all this we finally find
\begin{equation}
 \nu_{H^0(X,N)}=\nu_{H^0({\cal N})}-\nu_{H^0({\cal N}^*\otimes{\cal N})}+\nu_{H^1({\cal N}^*\otimes{\cal N})}+ \nu_{H^2(\wedge^2{\cal N}^*\otimes{\cal N})}
 =13-2+1+0=12\; .
\end{equation} \\[2mm]
\underline{\bf Downstairs Hodge numbers}: This leads to the downstairs Hodge numbers
\begin{eqnarray}
 h^{2,1}(\tilde{X})&=&\nu_{H^0(X,N)}- \nu_{H^0(X,S)}+\nu_{H^0(X,{\cal O}_X^{\oplus 6})}=12-6+2=8\\
 h^{1,1}(\tilde{X})&=&\eta(X)/8+h^{2,1}(\tilde{X})=-32/8+8=4\; .
\end{eqnarray}

\subsection{Example 5: A co-dimension four CICY with a $Z_2\times Z_8$ symmetry}
\underline{\bf Set-up}:  This example is for the CICY with number 6836, embedded in the ambient space ${\cal A}=\mathbb{P}^1\times \mathbb{P}^1\times \mathbb{P}^1\times \mathbb{P}^1\times \mathbb{P}^3$ and specified by the configuration matrix
\begin{equation}
X=\left[\begin{array}{l|llll}\mathbb{P}^1&0&0&0&2\\\mathbb{P}^1&0&2&0&0\\\mathbb{P}^1&0&0&2&0\\\mathbb{P}^1&2&0&0&0\\
\mathbb{P}^3&1&1&1&1\end{array}\right]^{5,37}_{-64}\; .
\end{equation}
The relevant ambient space bundles, besides $ {\cal O}_{\cal A}^{\oplus 5}$, are
\begin{eqnarray}
 {\mathfrak S}&=&{\cal O}_{\cal A}({\bf e}_1)^{\oplus 2}\oplus{\cal O}_{\cal A}({\bf e}_2)^{\oplus 2}\oplus{\cal O}_{\cal A}({\bf e}_3)^{\oplus 2}\oplus{\cal O}_{\cal A}({\bf e}_4)^{\oplus 2}\oplus{\cal O}_{\cal A}({\bf e}_5)^{\oplus 4}\\
 {\cal N}&=&{\cal N}_1\oplus{\cal N}_2\oplus{\cal N}_3\oplus{\cal N}_4\nonumber\\
 &=&{\cal O}_{\cal A}(0,0,0,2,1)\oplus{\cal O}_{\cal A}(0,2,0,0,1)\oplus{\cal O}_{\cal A}(0,0,2,0,1)\oplus{\cal O}_{\cal A}(2,0,0,0,1)
 \end{eqnarray}\\[-2mm]
\underline{\bf Cohomologies}:  Chasing through the Koszul sequence for $S={\mathfrak S}|_X$ it follows that
\begin{eqnarray}
 H^0(X,S)&\cong& H^0({\mathfrak S})\oplus H^1({\cal N}^*\otimes{\mathfrak S})\label{S1}\\\nonumber
 H^0({\mathfrak S})&=&\underbrace{H^0({\cal O}_{\cal A}({\bf e}_1))^{\oplus 2}}_{\text{4 dim.}}\oplus \underbrace{H^0({\cal O}_{\cal A}({\bf e}_2))^{\oplus 2}}_{\text{4 dim.}}\oplus \underbrace{H^0({\cal O}_{\cal A}({\bf e}_3))^{\oplus 2}}_{\text{4 dim.}}\oplus \underbrace{H^0({\cal O}_{\cal A}({\bf e}_4))^{\oplus 2}}_{\text{4 dim.}}\label{S2}\\[1pt]
& & \hspace{259pt}\oplus\underbrace{H^0({\cal O}_{\cal A}({\bf e}_5))^{\oplus 4}}_{\text{16 dim.}}\\[4pt]\nonumber
 H^1({\cal N}^*\otimes{\mathfrak S})&=&\left[H^1({\cal N}_1^*\otimes {\cal O}_{\cal A}({\bf e}_5))\oplus H^1({\cal N}_2^*\otimes {\cal O}_{\cal A}({\bf e}_5))\oplus H^1({\cal N}_3^*\otimes {\cal O}_{\cal A}({\bf e}_5))\oplus H^1({\cal N}_4^*\otimes {\cal O}_{\cal A}({\bf e}_5))\right]^{\oplus 4}\label{S3}\\\nonumber
 &=&\left[\underbrace{H^1({\cal O}_{\cal A}(0,0,0,-2,0))}_{\text{1 dim.}}\oplus \underbrace{H^1({\cal O}_{\cal A}(0,-2,0,0,0))}_{\text{1 dim.}}\oplus \underbrace{H^1({\cal O}_{\cal A}(0,0,-2,0,0))}_{\text{1 dim.}}\right.\\[-4pt]
& & \hspace{240pt}\left.\oplus \underbrace{H^1({\cal O}_{\cal A}(-2,0,0,0,0))}_{\text{1 dim.}}\right]^{\oplus 4}
\end{eqnarray} 
Altogether we have $h^0(X,S)=4\times 4+16+4\times 4=48$. Carrying out a similar discussion starting from the Koszul sequence for $N={\cal N}|_X$ we find
\begin{eqnarray}
 H^0(X,N)&\cong&\frac{H^0({\cal N})}{H^0({\cal N}^*\otimes{\cal N})}\oplus H^1({\cal N}^*\otimes{\cal N})\label{N1}\\
 H^0({\cal N})&=&\bigoplus_{a=1}^4 \underbrace{H^0({\cal N}_a)}_{\text{ 12 dim., each}}\\
 H^0({\cal N}^*\otimes{\cal N})&=&\bigoplus_{a=1}^4 H^0({\cal N}_a^*\otimes {\cal N}_a)=\bigoplus_{a=1}^4\underbrace{H^0({\cal O}_{\cal A})}_{\text{1 dim., each}}\\
H^1({\cal N}^*\otimes{\cal N})&=&\bigoplus_{a\neq b}\underbrace{H^1({\cal N}_a^*\otimes {\cal N}_b)}_{\text{12 terms, each 3 dim.}}\label{N4}
\end{eqnarray} 
This gives $h^0(X,N)=48-4+36=80$. Since $h^1(X,S)=0$ the kernel in Eq.~\eqref{h21form} vanishes.\\[8mm]
\underline{\bf Upstairs Hodge number}: From the above cohomologies we find
\begin{equation}
 h^{2,1}(X)=h^0(X,N)-h^0(X,S)+5=80-48+5=37
\end{equation}
as expected.\\[8mm]
\underline{\bf Symmetry}: We consider the symmetry with number 117 on this manifold, with group $G=\mathbb{Z}_2\times \mathbb{Z}_8$, $\mathbb{Z}_2$ generator $g_1$ and $\mathbb{Z}_8$ generator $g_2$. The action on the homogenous ambient space coordinate is given by
\begin{eqnarray}
\gamma(g_1)&=&{\rm diag}(1,-1,-1,1,1,-1,-1,1,1,-1,1,-1)\\[2mm]
\gamma(g_2)&=&\scriptsize\left(
\begin{array}{cccccccccccc}
 0 & 0 & 0 & 0 & 1 & 0 & 0 & 0 & 0 & 0 & 0 & 0 \\
 0 & 0 & 0 & 0 & 0 & 1 & 0 & 0 & 0 & 0 & 0 & 0 \\
 0 & 0 & 0 & 0 & 0 & 0 & 0 & 1 & 0 & 0 & 0 & 0 \\
 0 & 0 & 0 & 0 & 0 & 0 & -1 & 0 & 0 & 0 & 0 & 0 \\
 0 & 0 & -1 & 0 & 0 & 0 & 0 & 0 & 0 & 0 & 0 & 0 \\
 0 & 0 & 0 & 1 & 0 & 0 & 0 & 0 & 0 & 0 & 0 & 0 \\
 1 & 0 & 0 & 0 & 0 & 0 & 0 & 0 & 0 & 0 & 0 & 0 \\
 0 & 1 & 0 & 0 & 0 & 0 & 0 & 0 & 0 & 0 & 0 & 0 \\
 0 & 0 & 0 & 0 & 0 & 0 & 0 & 0 & 0 & i & 0 & 0 \\
 0 & 0 & 0 & 0 & 0 & 0 & 0 & 0 & 1 & 0 & 0 & 0 \\
 0 & 0 & 0 & 0 & 0 & 0 & 0 & 0 & 0 & 0 & 0 & -i \\
 0 & 0 & 0 & 0 & 0 & 0 & 0 & 0 & 0 & 0 & 1 & 0 \\
\end{array}
\right)\; ,
\end{eqnarray}
while the action on the defining polynomials is
\begin{equation}
\rho(g_1)=\left(
\begin{array}{cccc}
 1 & 0 & 0 & 0 \\
 0 & -1 & 0 & 0 \\
 0 & 0 & 1 & 0 \\
 0 & 0 & 0 & -1 \\
\end{array}
\right)
\qquad\qquad\rho(g_2)=\left(
\begin{array}{cccc}
 0 & 0 & 0 & 1 \\
 1 & 0 & 0 & 0 \\
 0 & 1 & 0 & 0 \\
 0 & 0 & -1 & 0 \\
\end{array}
\right)\; .
\end{equation}
The associated permutation representation reads
\begin{equation}
\pi(g_1)=\mathbf{1}_5\qquad\qquad
\pi(g_2)=\left(
\begin{array}{ccccc}
 0 & 0 & 1 & 0 & 0 \\
 0 & 0 & 0 & 1 & 0 \\
 0 & 1 & 0 & 0 & 0 \\
 1 & 0 & 0 & 0 & 0 \\
 0 & 0 & 0 & 0 & 1 \\
\end{array}
\right)\; .
\end{equation}\\[2mm]
\underline{\bf Singlets}: {\bf a)}  We begin by computing the number of $G$-singlets in $H^0(X,{\cal O}_X^{\oplus 5})$. Imposing $\pi(g_2){\bf c}={\bf c}$ on ${\bf c}=(c_1,\ldots ,c_5)^T$ immediately shows that there are two singlets so
\begin{equation}
 \nu_{H^0(X,{\cal O}_X^{\oplus 5})}=2\; .
\end{equation} 
{\bf b)} For $H^0(X,S)$ we have to go through the various pieces in Eqs.~\eqref{S1}--\eqref{S3}. Setting up a 12-dimensional vector ${\bf l}$ of general linear polynomials, representing $H^0({\mathfrak S})$ and imposing $\gamma(g){\bf l}(\gamma (g)^{-1}{\bf x})={\bf l}({\bf x})$ for $g=g_1,g_2$ shows that
\begin{equation}
 \nu_{H^0({\mathfrak S})}=3\; .
\end{equation} 
The other relevant cohomology, $H^1({\cal N}^*\otimes{\mathfrak S})$, is represented by a vector
\begin{equation}
{\bf v}({\bf x})=\left(\frac{a_1}{t_0t_1},\frac{a_2}{y_0y_1},\frac{a_3}{z_0z_1},\frac{a_4}{x_0x_1}\right)^T
\end{equation}
where $((x_0,x_1),(y_0,y_1),(z_0,z_1),(t_0,t_1),(u_0,u_1,u_2,u_3))$ are the projective ambient space coordinates and $a_i$ are arbitrary constants. Imposing $\rho(g)^\dagger{\bf v}(\gamma(g)^{-1}({\bf x}))={\bf v}({\bf x})$ shows that
\begin{equation}
 \nu_{H^1({\cal N}^2\otimes{\mathfrak S})}=0\; .
\end{equation}
Combining these results we find that
\begin{equation}
 \nu_{H^0(X,S)}= \nu_{H^0({\mathfrak S})}+\nu_{H^1({\cal N}^*\otimes{\mathfrak S})}=3+0=3\; .
\end{equation} 
{\bf c)} For the final piece, $H^0(X,N)$, we have to consider the various contributions in Eqs.~\eqref{N1}--\eqref{N4}, beginning with $H^0({\cal N})$. We set up a four dimensional vector ${\bf v}$ of general polynomials describing $H^0({\cal N})$ and imposing $\rho(g){\bf v}(\gamma(g)^{-1}({\bf x}))={\bf v}({\bf x})$ shows that
\begin{equation}
 \nu_{H^0({\cal N})}=4\; .
\end{equation} 
The cohomology $H^0({\cal N}^*\otimes {\cal N})$ can be represented by a diagonal matrix $M={\rm diag}(a_1,a_2,a_3,a_4)$ and imposing the constraints $\rho(g)M\rho(g)^\dagger=M$ shows that
\begin{equation}
 \nu_{H^0({\cal N}^*\otimes {\cal N})}=1\; .
\end{equation}
Finally, $H^1({\cal N}^*\otimes {\cal N})$ can be represented by a polynomial $4\times 4$ matrix $M({\bf x})$ which has zero diagonal entries and $12$ general polynomials of the appropriate degree, as in Eq.~\eqref{N4}, in the off-diagonal entries. Imposing $\rho(g)M(\gamma(g)^{-1}{\bf x})\rho(g)^\dagger=M({\bf x})$ leads to
\begin{equation}
 \nu_{H^1({\cal N}^*\otimes {\cal N})}=2\; .
\end{equation}
Combining these three results we have
\begin{equation}
 \nu_{H^0(X,N)}=  \nu_{H^0({\cal N})}- \nu_{H^0({\cal N}^*\otimes {\cal N})}+ \nu_{H^1({\cal N}^*\otimes {\cal N})}=4-1+2=5\; .
\end{equation}\\[2mm] 
\underline{\bf Downstairs Hodge numbers}: For the Hodge numbers these results imply 
\begin{eqnarray}
 h^{2,1}(\tilde{X})&=&\nu_{H^0(X,N)}-\nu_{H^0(X,S)}+\nu_{H^0(X,{\cal O}_X^{\oplus 5})}=5-3+2=4\\
 h^{1,1}(\tilde{X})&=&\eta(X)/32+ h^{2,1}(\tilde{X})=-64/32+4=2
\end{eqnarray}

\section{Conclusion}
The present work concludes and completes the series of papers \cite{Candelas:2008wb, Candelas:2010ve, Candelas:2015amz} whose purpose is the computation of  Hodge numbers for smooth quotients of CICY manifolds.  Our results are summarised in Figure~\ref{TipHodgePlot}. This figure represents the tip of the Hodge plot of all Calabi-Yau manifolds presently known, highlighting CICY quotients and the new results obtained in the present paper. 
\begin{figure}[h!]
\begin{center}
\includegraphics[width=15.1cm]{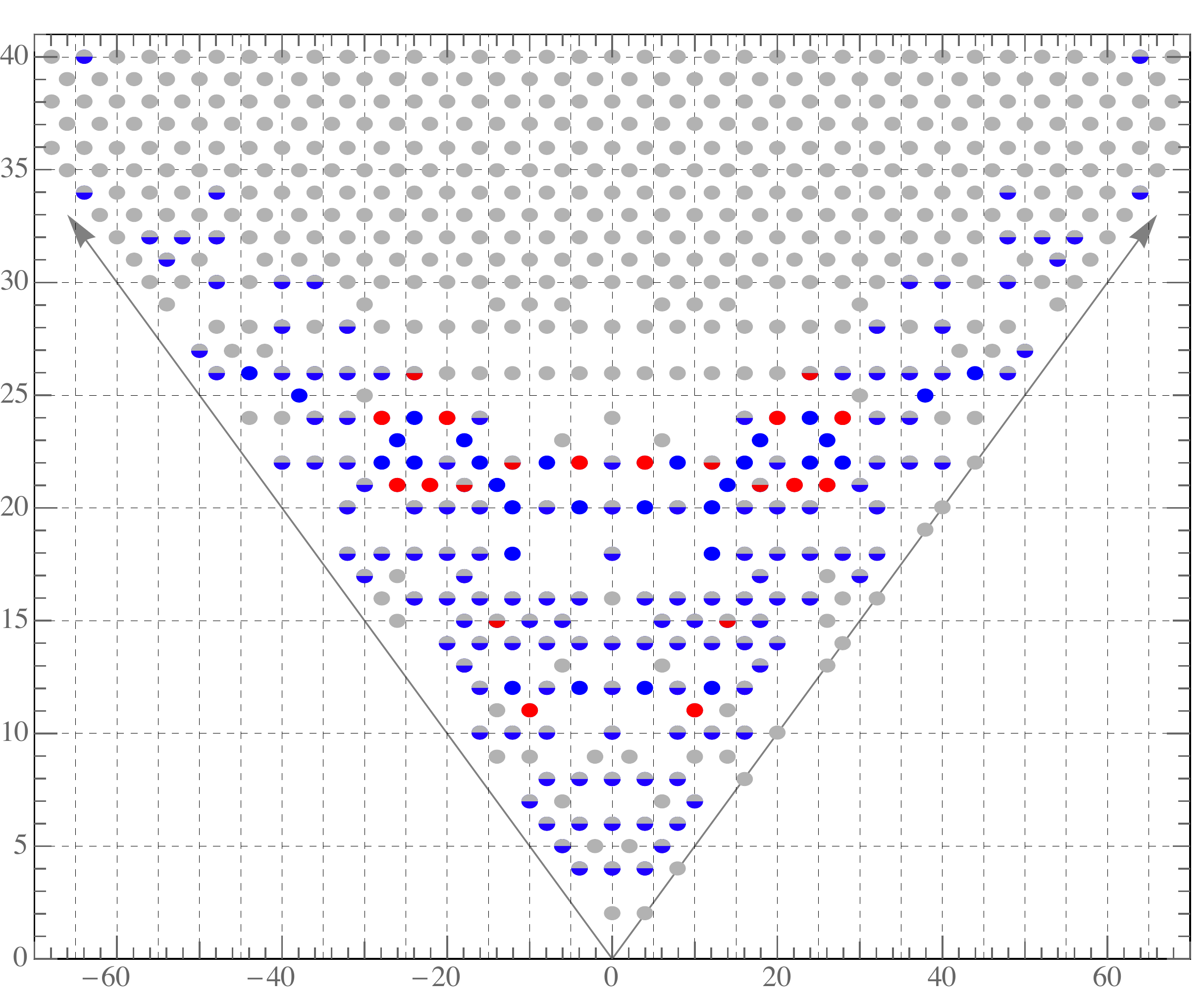}
%\vskip3pt
\capt{5.8in}{TipHodgePlot}{The tip of the Hodge number plot for all the Calabi-Yau three-folds that we know. Coloured points correspond to CICY quotients together their mirrors; the red points correspond to quotients studied only in the present paper. Monochrome points indicate quotients whose Hodge numbers fall onto sites previously unoccupied, while the multicoloured points correspond to multiply occupied sites.}\label{fig:Hodge}
\end{center}
\end{figure}

The detailed results of our computation, that is, the Hodge numbers for all smooth CICY quotients, can be found in the Appendix. These numbers are important for string model building on CICY quotients, particularly in the context of the heterotic compactifications. At the same time, CICY quotients provide examples of manifolds with small Hodge numbers - such examples are rare in the space of known Calabi-Yau manifolds. To our knowledge, the present method for computing Hodge numbers of CICY quotients based on equivariant cohomology has not been systematically applied before. The methods for computing equivariant cohomology developed in this context will be crucial for the analysis of the large dataset of heterotic models in Ref.~\cite{Anderson:2013xka}.

\section*{Acknowledgements}
We would like to thank Philip Candelas and Challenger Mishra for very helpful discussions. 
A.C.~would like to thank the Physics department at the University of Oxford for hospitality, where part of this work has been carried out.
The work of J.G.~is supported by NSF grant PHY-1417316.
A.L.~is partially supported by the EPSRC network grant EP/N007158/1 and by the STFC grant~ST/L000474/1.

\newpage
\begin{appendix}
\section{Table of Hodge numbers}\label{appendix}
In this appendix we cover all CICYs which appear in the original list~\cite{Candelas:1987kf, Candelas:1987du} and which have freely-acting symmetries according to the classification by Braun in Ref.~\cite{Braun:2010vc}. The Hodge numbers of all resulting quotients are listed in the table below. We have marked in red the Hodge numbers which have not been computed elsewhere, to our knowledge. For convenience, the underlying data can be obtained from the website~\cite{CicySymm}. This data also contains the explicit CICY configuration matrices and the representations $\gamma$, $\rho$ of the freely-acting symmetries which have not been included in the table below in order to keep the size manageable. Instead, the table refers to the relevant CICYs and their symmetries by identifier numbers (in the first and fourth column, respectively), which directly refer to the data at~\cite{CicySymm}. Further, the second column lists the upstairs Hodge numbers, the third column the symmetry group $G$ and the last column the downstairs Hodge numbers.
\end{appendix}

\begin{center}
\begin{longtable}{|c|c|>{~~}c<{~~}|>{\quad}c<{\quad}|>{~}c<{~}|}

\hline \multicolumn{1}{|c|}{\str\textbf{~~CICY \#~~}}&  \multicolumn{1}{|c|}{\textbf{$ ~( h^{1,1}(X),\, h^{2,1}(X))~$}} & \multicolumn{1}{|c|}{\str\textbf{$G$}}&\multicolumn{1}{|c|}{\str\textbf{~Symm \#~}}& \multicolumn{1}{|c|}{$ ( h^{1,1}(X/G),\, h^{2,1}(X/G))$} \\ \hline 
\endfirsthead

\hline 
\str\textbf{CICY \#} &
$(h^{1,1}(X),\, h^{2,1}(X))$ &
$G$&
\str\textbf{~~Symm \#~~}&
$ ( h^{1,1}(X/G),\, h^{2,1}(X/G))$ \\ \hline 
\endhead

\hline\hline \multicolumn{5}{|r|}{{\str Continued on next page}} \\ \hline
\endfoot

\hline\hline\multicolumn{5}{|c|}{\str}\\ \hline
\endlastfoot

\hline\hline

\str$4$&$(15,15)$&$\IZ_2$&$1-16$&$(9,9)$\\  \hline
\str$5$&$(15,15)$&$\IZ_2$&$1-32$&$(9,9)$\\  \hline
\str$6$&$(15,15)$&$\IZ_2$&$1-32$&$(9,9)$\\
\strr & & $\IZ_3$&$33$&$(7,7)$\\
\strr & & $\IZ_6$&$34-41$&$(3,3)$\\  \hline
\str$14$&$(19,19)$&$\IZ_3$&$1-3$&$(7,7)$\\
\strr & & $\IZ_3 \times \IZ_3$&$4-39$&$(3,3)$\\  \hline
\str$15$&$(15,15)$&$\IZ_2$&$1-4$&$(9,9)$\\
\strr & & $\IZ_2 \times \IZ_2$&$5-20$&$(6,6)$\\  \hline
\str$18$&$(19,19)$&$\IZ_3$&$1-3$&$(7,7)$\\  \hline
\str$19$&$(19,19)$&$\IZ_2$&$1-3$&$(11,11)$\\
\strr & & $\IZ_4$&$4-9$&$(5,5)$\\
\strr & & $\IZ_2 \times \IZ_2$&$10-16$&$(7,7)$\\
\strr & & $\IZ_8$&$17$&$(3,3)$\\
\strr & & $\IZ_4 \times \IZ_2$&$18-20$&$(4,4)$\\  
\strr & & $\IQ_8$ & $21-29$ & $(3,3)$\\ \hline
\str$20$&$(19,19)$&$\IZ_2$&$1-2$&$(11,11)$\\
\strr & & $\IZ_4$&$3-7$&$(5,5)$\\
\strr & & $\IZ_2 \times \IZ_2$&$8-14$&$(7,7)$\\  \hline
\str$21$&$(19,19)$&$\IZ_2$&$1-3$&$(11,11)$\\
\strr & & $\IZ_4$&$4-9$&$(5,5)$\\
\strr & & $\IZ_2 \times \IZ_2$&$10-16$&$(7,7)$\\
\strr & & $\IZ_8$&$17$&$(3,3)$\\
\strr & & $\IZ_4 \times \IZ_2$&$18-28$&$(3,3)$\\
\strr & & $\IZ_4 \times \IZ_2$&$29-31$&$(4,4)$\\
\strr & & $\IQ_8$ & $32-34$ & $(3,3)$\\ 
\strr & & $\IZ_4 \times \IZ_4$&$35-37$&$(2,2)$\\
\strr & & $\IZ_4 \rtimes \IZ_4$ & $38-40$ & $(2,2)$\\
\strr & & $\IZ_8 \times \IZ_2$&$41-43$&$(2,2)$\\  
\strr & & $\IZ_8 \rtimes \IZ_2$ & $44-46$ & $(2,2)$\\
\str & & $\IZ_2 	\times \IQ_8$ & $47-53$ & $(2,2)$\\ \hline
\str$22$&$(15,15)$&$\IZ_2$&$1-9$&$(9,9)$\\
\strr & & $\IZ_2 \times \IZ_2$&$10-15$&$(7,7)$\\  \hline
\str$26$&$(19,19)$&$\IZ_3$&$1-3$&$(7,7)$\\  \hline
\str$27$&$(19,19)$&\tr{$\IZ_2$}&\tr{$1-2$}&\tr{$(11,11)$}\\  \hline
\str$28$&$(19,19)$&\tr{$\IZ_2$}&\tr{$1-3$}&\tr{$(11,11)$}\\  \hline
\str$30$&$(19,19)$&$\IZ_2$&$1-3$&$(11,11)$\\
\strr & & $\IZ_4$&$4$&$(6,6)$\\  \hline
\str$90$&$(13,17)$&$\IZ_2$&$1$&$(9,11)$\\  \hline
\str$95$&$(16,20)$&\tr{$\IZ_2$}&\tr{$1$}&\tr{$(10,12)$}\\  \hline
\str$111$&$(14,18)$&\tr{$\IZ_2$}&\tr{$1$}&\tr{$(9,11)$}\\  \hline
\str$242$&$(12,18)$&$\IZ_3$&$1$&$(6,8)$\\  \hline
\str$261$&$(11,19)$&$\IZ_2$&$1$&$(8,12)$\\  \hline
\str$343$&$(11,19)$&$\IZ_2$&$1$&$(8,12)$\\  \hline
\str$376$&$(11,19)$&$\IZ_2$&$1$&$(8,12)$\\  \hline
\str$379$&$(11,19)$&$\IZ_2$&$1$&$(8,12)$\\  \hline
\str$381$&$(14,22)$&\tr{$\IZ_2$}&\tr{$1$}&\tr{$(9,13)$}\\  \hline
\str$382$&$(14,22)$&\tr{$\IZ_2$}&\tr{$1-3$}&\tr{$(9,13)$}\\  \hline
\str$397$&$(12,20)$&\tr{$\IZ_2$}&\tr{$1$}&\tr{$(8,12)$}\\  \hline
\str$399$&$(12,20)$&\tr{$\IZ_2$}&\tr{$1-8$}&\tr{$(8,12)$}\\  \hline
\str$400$&$(12,20)$&\tr{$\IZ_2$}&\tr{$1-8$}&\tr{$(8,12)$}\\  \hline
\str$401$&$(12,20)$&\tr{$\IZ_2$}&\tr{$1-16$}&\tr{$(8,12)$}\\  \hline
\str$402$&$(12,20)$&\tr{$\IZ_2$}&\tr{$1-16$}&\tr{$(8,12)$}\\  \hline
\str$480$&$(13,21)$&$\IZ_2$&$1-21$&$(9,13)$\\
\strr & & $\IZ_4$&$22-26$&$(5,7)$\\
\strr & & $\IZ_2 \times \IZ_2$&$27-394$&$(7,9)$\\  \hline
\str$536$&$(14,23)$&$\IZ_3$&$1$&$(6,9)$\\  \hline
\str$1144$&$(11,23)$&\tr{$\IZ_2$}&\tr{$1-20$}&\tr{$(8,14)$}\\  \hline
\str$1215$&$(9,21)$&$\IZ_3$&$1$&$(5,9)$\\  \hline
\str$1257$&$(12,24)$&\tr{$\IZ_2$}&\tr{$1-2$}&\tr{$(8,14)$}\\  \hline
\str$1262$&$(9,21)$&$\IZ_2$&$1$&$(7,13)$\\  \hline
\str$1268$&$(12,24)$&\tr{$\IZ_2$}&\tr{$1$}&\tr{$(8,14)$}\\  \hline
\str$1270$&$(10,22)$&\tr{$\IZ_2$}&\tr{$1$}&\tr{$(7,13)$}\\  \hline
\str$1295$&$(11,23)$&\tr{$\IZ_2$}&\tr{$1-5$}&\tr{$(8,14)$}\\  \hline
\str$1298$&$(10,22)$&\tr{$\IZ_2$}&\tr{$1$}&\tr{$(7,13)$}\\  \hline
\str$1306$&$(9,21)$&$\IZ_3$&$1$&$(5,9)$\\  \hline
\str$1441$&$(10,22)$&\tr{$\IZ_2$}&\tr{$1$}&\tr{$(7,13)$}\\  \hline
\str$1701$&$(9,23)$&$\IZ_2$&$1$&$(7,14)$\\  \hline
\str$2104$&$(11,26)$&$\IZ_3$&$1$&$(5,10)$\\  \hline
\str$2357$&$(9,25)$&$\IZ_2$&$1-10$&$(7,15)$\\
\strr & & $\IZ_2 \times \IZ_2$&$11-98$&$(6,10)$\\  \hline
\str$2360$&$(10,26)$&\tr{$\IZ_2$}&\tr{$1$}&\tr{$(7,15)$}\\  \hline
\str$2374$&$(9,25)$&\tr{$\IZ_2$}&\tr{$1$}&\tr{$(6,14)$}\\  \hline
\str$2383$&$(9,25)$&\tr{$\IZ_2$}&\tr{$1-5$}&\tr{$(7,15)$}\\  \hline
\str$2533$&$(10,26)$&\tr{$\IZ_2$}&\tr{$1-4$}&\tr{$(7,15)$}\\  \hline
\str$2534$&$(9,25)$&$\IZ_2$&$1-5$&$(7,15)$\\
\strr & & $\IZ_2 \times \IZ_2$&$6-27$&$(6,10)$\\  \hline
\str$2535$&$(9,25)$&\tr{$\IZ_2$}&\tr{$1-8$}&\tr{$(7,15)$}\\  \hline
\str$2536$&$(9,25)$&\tr{$\IZ_2$}&\tr{$1-8$}&\tr{$(7,15)$}\\  \hline
\str$2543$&$(9,25)$&\tr{$\IZ_2$}&\tr{$1-4$}&\tr{$(7,15)$}\\  \hline
\str$2544$&$(7,23)$&$\IZ_2$&$1$&$(6,14)$\\  \hline
\str$2564$&$(12,28)$&$\IZ_2$&$1$&$(8,16)$\\
\strr & & $\IZ_4$&$2-3$&$(4,8)$\\
\strr & & $\IZ_2 \times \IZ_2$&$4$&$(6,10)$\\
\strr & & $\IZ_8$&$5$&$(2,4)$\\
\strr & & $\IZ_4 \times \IZ_2$&$6$&$(3,5)$\\  
\strr & & $\IQ_8$ & $7-9$ & $(2,4)$\\ \hline
\str$2565$&$(10,26)$&\tr{$\IZ_2$}&\tr{$1$}&\tr{$(7,15)$}\\  \hline
\str$2566$&$(12,28)$&$\IZ_2$&$1-3$&$(8,16)$\\
\strr & & $\IZ_2 \times \IZ_2$&$4-10$&$(6,10)$\\  \hline
\str$2568$&$(12,28)$&$\IZ_2$&$1-7$&$(8,16)$\\
\strr & & $\IZ_4$&$8$&$(4,8)$\\
\strr & & $\IZ_2 \times \IZ_2$&$9-40$&$(6,10)$\\
\strr & & $\IZ_4 \times \IZ_2$&$41-42$&$(3,5)$\\  \hline
\str$2570$&$(8,24)$&\tr{$\IZ_2$}&\tr{$1$}&\tr{$(6,14)$}\\  \hline
\str$2572$&$(9,25)$&$\IZ_2$&$1$&$(7,15)$\\
\strr & & $\IZ_4$&$2$&$(4,8)$\\  \hline
\str$2639$&$(9,25)$&$\IZ_2$&$1-16$&$(7,15)$\\
\strr & & $\IZ_4$&$17-20$&$(4,8)$\\  \hline
\str$2640$&$(9,25)$&$\IZ_2$&$1-5$&$(7,15)$\\
\strr & & $\IZ_2 \times \IZ_2$&$6-27$&$(6,10)$\\  \hline
\str$2654$&$(9,25)$&\tr{$\IZ_2$}&\tr{$1-8$}&\tr{$(7,15)$}\\  \hline
\str$2655$&$(10,26)$&\tr{$\IZ_2$}&\tr{$1-8$}&\tr{$(7,15)$}\\  \hline
\str$2660$&$(9,25)$&\tr{$\IZ_2$}&\tr{$1-10$}&\tr{$(7,15)$}\\  \hline
\str$2839$&$(9,25)$&\tr{$\IZ_2$}&\tr{$1-5$}&\tr{$(7,15)$}\\  \hline
\str$3381$&$(9,27)$&$\IZ_2$&$1$&$(7,16)$\\  \hline
\str$3388$&$(11,29)$&$\IZ_3$&$1$&$(5,11)$\\  \hline
\str$3406$&$(11,29)$&$\IZ_3$&$1-3$&$(5,11)$\\  \hline
\str$3413$&$(6,24)$&$\IZ_3$&$1$&$(4,10)$\\  \hline
\str$3496$&$(8,26)$&\tr{$\IZ_2$}&\tr{$1$}&\tr{$(6,15)$}\\  \hline
\str$3620$&$(9,27)$&$\IZ_3$&$1$&$(5,11)$\\  \hline
\str$3929$&$(7,27)$&$\IZ_2$&$1$&$(6,16)$\\  \hline
\str$3939$&$(8,28)$&\tr{$\IZ_2$}&\tr{$1-5$}&\tr{$(6,16)$}\\  \hline
\str$4071$&$(7,27)$&\tr{$\IZ_2$}&\tr{$1-2$}&\tr{$(6,16)$}\\  \hline
\str$4078$&$(7,27)$&\tr{$\IZ_2$}&\tr{$1-4$}&\tr{$(6,16)$}	\\  \hline
\str$4086$&$(8,28)$&\tr{$\IZ_2$}&\tr{$1$}&\tr{$(6,16)$}\\  \hline
\str$4108$&$(7,27)$&$\IZ_2$&$1$&$(6,16)$\\  \hline
\str$4109$&$(6,26)$&$\IZ_2$&$1$&$(5,15)$\\  \hline
\str$4185$&$(7,27)$&\tr{$\IZ_2$}&\tr{$1-8$}&\tr{$(6,16)$}\\  \hline
\str$4197$&$(9,29)$&\tr{$\IZ_2$}&\tr{$1$}&\tr{$(6,16)$}\\  \hline
\str$4216$&$(8,28)$&\tr{$\IZ_2$}&\tr{$1$}&\tr{$(5,15)$}\\  \hline
\str$4224$&$(10,30)$&\tr{$\IZ_2$}&\tr{$1-8$}&\tr{$(7,17)$}\\  \hline
\str$4227$&$(8,28)$&\tr{$\IZ_2$}&\tr{$1-5$}&\tr{$(6,16)$}\\  \hline
\str$4335$&$(7,27)$&$\IZ_2$&$1$&$(6,16)$\\
\strr & & $\IZ_5$&$2$&$(3,7)$\\
\strr & & $\IZ_{10}$&$3$&$(2,4)$\\  \hline
\str$4415$&$(8,29)$&$\IZ_3$&$1$&$(4,11)$\\  \hline
\str$4738$&$(7,29)$&\tr{$\IZ_2$}&\tr{$1$}&\tr{$(5,16)$}\\  \hline
\str$5141$&$(7,31)$&\tr{$\IZ_2$}&\tr{$1-4$}&\tr{$(6,18)$}\\  \hline
\str$5248$&$(7,31)$&\tr{$\IZ_2$}&\tr{$1-4$}&\tr{$(6,18)$}\\  \hline
\str$5254$&$(10,34)$&\tr{$\IZ_2$}&\tr{$1-2$}&\tr{$(7,19)$}\\  \hline
\str$5256$&$(5,29)$&$\IZ_2$&$1-2$&$(5,17)$\\
\strr & & $\IZ_2 \times \IZ_2$&$3-6$&$(5,11)$\\  \hline
\str$5257$&$(10,34)$&\tr{$\IZ_2$}&\tr{$1-6$}&\tr{$(7,19)$}\\  \hline
\str$5259$&$(7,31)$&\tr{$\IZ_2$}&\tr{$1-8$}&\tr{$(6,18)$}\\  \hline
\str$5273$&$(6,30)$&$\IZ_2$&$1$&$(5,17)$\\  \hline
\str$5277$&$(7,31)$&\tr{$\IZ_2$}&\tr{$1$}&\tr{$(6,18)$}\\  \hline
\str$5300$&$(10,34)$&\tr{$\IZ_2$}&\tr{$1$}&\tr{$(7,19)$}\\  \hline
\str$5301$&$(5,29)$&$\IZ_2$&$1$&$(5,17)$\\
\strr & & $\IZ_4$&$2$&$(3,9)$\\
\strr & & $\IZ_2 \times \IZ_2$&$3$&$(5,11)$\\  \hline
\str$5302$&$(6,30)$&$\IZ_2$&$1-4$&$(6,18)$\\
\strr & & $\IZ_2 \times \IZ_2$&$5-20$&$(6,12)$\\  \hline
\str$5306$&$(10,34)$&\tr{$\IZ_2$}&\tr{$1-3$}&\tr{$(7,19)$}\\  \hline
\str$5310$&$(7,31)$&\tr{$\IZ_2$}&\tr{$1$}&\tr{$(5,17)$}\\  \hline
\str$5311$&$(7,31)$&\tr{$\IZ_2$}&\tr{$1-2$}&\tr{$(6,18)$}\\  \hline
\str$5406$&$(7,31)$&\tr{$\IZ_2$}&\tr{$1-4$}&\tr{$(6,18)$}\\  \hline
\str$5421$&$(8,32)$&$\IZ_2$&$1-5$&$(6,18)$\\
\strr & & $\IZ_2 \times \IZ_2$&$6-27$&$(5,11)$\\  \hline
\str$5423$&$(7,31)$&$\IZ_2$&$1$&$(6,18)$\\  \hline
\str$5425$&$(6,30)$&$\IZ_2$&$1$&$(5,17)$\\  \hline
\str$5449$&$(7,31)$&\tr{$\IZ_2$}&\tr{$1-5$}&\tr{$(5,17)$}\\  \hline
\str$5452$&$(5,29)$&$\IZ_2$&$1-4$&$(5,17)$\\
\strr & & $\IZ_4$&$5-6$&$(3,9)$\\
\strr & & $\IZ_2 \times \IZ_2$&$7-22$&$(5,11)$\\  \hline
\str$5826$&$(6,32)$&\tr{$\IZ_2$}&\tr{$1$}&\tr{$(4,17)$}\\  \hline
\str$5958$&$(6,32)$&$\IZ_2$&$1$&$(5,18)$\\  \hline
\str$5967$&$(6,33)$&$\IZ_3$&$1$&$(4,13)$\\  \hline
\str$5982$&$(6,33)$&$\IZ_3$&$1$&$(4,13)$\\  \hline
\str$6021$&$(8,35)$&$\IZ_3$&$1$&$(4,13)$\\  \hline
\str$6024$&$(5,32)$&$\IZ_3$&$1$&$(3,12)$\\  \hline
\str$6173$&$(7,35)$&$\IZ_2$&$1$&$(6,20)$\\  \hline
\str$6178$&$(7,35)$&\tr{$\IZ_2$}&\tr{$1-4$}&\tr{$(6,20)$}\\  \hline
\str$6187$&$(6,34)$&\tr{$\IZ_2$}&\tr{$1-4$}&\tr{$(5,19)$}\\  \hline
\str$6201$&$(6,34)$&\tr{$\IZ_2$}&\tr{$1-4$}&\tr{$(5,19)$}\\  \hline
\str$6202$&$(6,34)$&\tr{$\IZ_2$}&\tr{$1$}&\tr{$(5,19)$}\\  \hline
\str$6204$&$(5,33)$&$\IZ_2$&$1$&$(4,18)$\\  \hline
\str$6225$&$(5,33)$&$\IZ_2$&$1$&$(4,18)$\\  \hline
\str$6229$&$(6,34)$&\tr{$\IZ_2$}&\tr{$1$}&\tr{$(5,19)$}\\  \hline
\str$6231$&$(6,34)$&\tr{$\IZ_2$}&\tr{$1$}&\tr{$(5,19)$}\\  \hline
\str$6281$&$(6,34)$&\tr{$\IZ_2$}&\tr{$1-4$}&\tr{$(5,19)$}\\  \hline
\str$6502$&$(7,37)$&$\IZ_3$&$1$&$(3,13)$\\  \hline
\str$6655$&$(6,36)$&$\IZ_5$&$1$&$(2,8)$\\  \hline
\str$6715$&$(5,37)$&$\IZ_2$&$1$&$(5,21)$\\
\strr & & $\IZ_2 \times \IZ_2$&$2$&$(5,13)$\\  \hline
\str$6724$&$(5,37)$&$\IZ_2$&$1$&$(4,20)$\\  \hline
\str$6732$&$(5,37)$&$\IZ_2$&$1-2$&$(5,21)$\\  \hline
\str$6738$&$(6,38)$&$\IZ_2$&$1$&$(5,21)$\\  \hline
\str$6770$&$(5,37)$&$\IZ_2$&$1-2$&$(5,21)$\\  \hline
\str$6777$&$(5,37)$&$\IZ_2$&$1-4$&$(5,21)$\\  \hline
\str$6780$&$(5,37)$&\tr{$\IZ_2$}&\tr{$1-4$}&\tr{$(4,20)$}\\  \hline
\str$6784$&$(4,36)$&$\IZ_2$&$1-2$&$(4,20)$\\
\strr & & $\IZ_2 \times \IZ_2$&$3-6$&$(4,12)$\\  \hline
\str$6785$&$(8,40)$&\tr{$\IZ_2$}&\tr{$1-4$}&\tr{$(6,22)$}\\  \hline
\str$6788$&$(5,37)$&$\IZ_2$&$1-3$&$(5,21)$\\
\strr & & $\IZ_2 \times \IZ_2$&$4-12$&$(5,13)$\\  \hline
\str$6802$&$(5,37)$&$\IZ_2$&$1$&$(5,21)$\\  \hline
\str$6804$&$(5,37)$&$\IZ_2$&$1$&$(4,20)$\\  \hline
\str$6826$&$(8,40)$&$\IZ_2$&$1$&$(6,22)$\\
\strr & & $\IZ_4$&$2$&$(3,11)$\\
\strr & & $\IZ_2 \times \IZ_2$&$3$&$(5,13)$\\  \hline
\str$6828$&$(4,36)$&$\IZ_2$&$1$&$(4,20)$\\
\strr & & $\IZ_2 \times \IZ_2$&$2$&$(4,12)$\\  \hline
\str$6829$&$(8,40)$&$\IZ_2$&$1-3$&$(6,22)$\\
\strr & & $\IZ_2 \times \IZ_2$&$4-10$&$(5,13)$\\  \hline
\str$6830$&$(5,37)$&\tr{$\IZ_2$}&\tr{$1$}&\tr{$(4,20)$}\\  \hline
\str$6831$&$(4,36)$&$\IZ_2$&$1$&$(3,19)$\\  \hline
\str$6834$&$(5,37)$&$\IZ_2$&$1-2$&$(5,21)$\\  \hline
\str$6836$&$(5,37)$&$\IZ_2$&$1-11$&$(5,21)$\\
\strr & & $\IZ_4$&$12-14$&$(3,11)$\\
\strr & & $\IZ_2 \times \IZ_2$&$15-92$&$(5,13)$\\
\strr & & $\IZ_8$&$93$&$(2,6)$\\
\strr & & $\IZ_4 \times \IZ_2$&$94-111$&$(3,7)$\\
\strr & & $\IQ_8$ & $112-113$ & $(2,6)$\\
\strr & & $\IZ_4 \rtimes \IZ_4$ & $114-115$ & $(2,4)$\\
\strr & & $\IZ_8 \times \IZ_2$&$116-117$&$(2,4)$\\  \hline
\str$6890$&$(5,37)$&$\IZ_2$&$1-2$&$(5,21)$\\  \hline
\str$6896$&$(5,37)$&$\IZ_2$&$1$&$(5,21)$\\  \hline
\str$6927$&$(5,37)$&$\IZ_2$&$1-2$&$(5,21)$\\
\strr & & $\IZ_4$&$3$&$(3,11)$\\
\strr & & $\IZ_2 \times \IZ_2$&$4-5$&$(5,13)$\\
\strr & & $\IZ_4 \times \IZ_2$&$6-8$&$(3,7)$\\  \hline
\str$6947$&$(5,37)$&$\IZ_2$&$1$&$(5,21)$\\
\strr & & $\IZ_4$&$2$&$(3,11)$\\
\strr & & $\IZ_2 \times \IZ_2$&$3$&$(5,13)$\\
\strr & & $\IZ_8$&$4$&$(2,6)$\\
\strr & & $\IZ_4 \times \IZ_2$&$5-6$&$(3,7)$\\
\strr & & $\IQ_8$ & $7$ & $(2,6)$\\
\strr & & $\IZ_4 \rtimes \IZ_4$ & $8$ & $(2,4)$\\
\strr & & $\IZ_8 \times \IZ_2$& $9$ &$(2,4)$\\  \hline
\str$7204$&$(4,40)$&$\IZ_2$&$1-2$&$(4,22)$\\  \hline
\str$7206$&$(8,44)$&$\IZ_2$&$1-2$&$(6,24)$\\
\strr & & $\IZ_3$&$3-5$&$(4,16)$\\
\strr & & $\IZ_6$&$6-11$&$(2,8)$\\  \hline
\str$7218$&$(4,40)$&$\IZ_2$&$1$&$(4,22)$\\  \hline
\str$7240$&$(3,39)$&$\IZ_3$&$1$&$(3,15)$\\
\strr & & $\IZ_3 \times \IZ_3$&$2-3$&$(3,7)$\\  \hline
\str$7241$&$(4,40)$&$\IZ_2$&$1$&$(4,22)$\\  \hline
\str$7245$&$(4,40)$&$\IZ_2$&$1$&$(3,21)$\\  \hline
\str$7246$&$(8,44)$&$\IZ_2$&$1$&$(6,24)$\\
\strr & & $\IZ_3$&$2-10$&$(4,16)$\\
\strr & & $\IZ_4$&$11$&$(3,12)$\\
\strr & & $\IZ_6$&$12-20$&$(2,8)$\\
\strr & & $\IZ_3 \rtimes \IZ_4$ & $21-23$ & $(1,4)$\\
\strr & & $\IZ_{12}$&$24-26$&$(1,4)$\\  \hline
\str$7247$&$(4,40)$&$\IZ_3$&$1$&$(2,14)$\\  \hline
\str$7270$&$(4,40)$&$\IZ_2$&$1-2$&$(4,22)$\\  \hline
\str$7279$&$(5,41)$&$\IZ_2$&$1$&$(4,22)$\\  \hline
\str$7300$&$(8,44)$&$\IZ_2$&$1-4$&$(6,24)$\\
\strr & & $\IZ_3$&$5$&$(4,16)$\\
\strr & & $\IZ_4$&$6-7$&$(3,12)$\\
\strr & & $\IZ_6$&$8-11$&$(2,8)$\\
\strr & & $\IZ_3 \rtimes \IZ_4$ & $12-13$ & $(1,4)$\\
\strr & & $\IZ_{12}$&$14-15$&$(1,4)$\\  \hline
\str$7403$&$(4,42)$&$\IZ_2$&$1$&$(3,22)$\\  \hline
\str$7435$&$(4,44)$&$\IZ_2$&$1$&$(4,24)$\\
\strr & & $\IZ_2 \times \IZ_2$&$2$&$(4,14)$\\  \hline
\str$7447$&$(5,45)$&$\IZ_2$&$1$&$(5,25)$\\
\strr & & $\IZ_2 \times \IZ_2$&$2$&$(5,15)$\\
\strr & & $\IZ_5$&$3$&$(1,9)$\\
\strr & & $\IZ_{10}$&$4$&$(1,5)$\\
\strr & & $\IZ_{10} \times \IZ_2$&$5$&$(1,3)$\\  \hline
\str$7450$&$(3,43)$&$\IZ_2$&$1-2$&$(3,23)$\\  \hline
\str$7462$&$(4,44)$&$\IZ_2$&$1-2$&$(4,24)$\\
\strr & & $\IZ_2 \times \IZ_2$&$3-4$&$(4,14)$\\  \hline
\str$7468$&$(4,44)$&$\IZ_2$&$1$&$(3,23)$\\  \hline
\str$7481$&$(3,43)$&$\IZ_2$&$1$&$(3,23)$\\  \hline
\str$7484$&$(3,43)$&$\IZ_2$&$1$&$(3,23)$\\
\strr & & $\IZ_4$&$2$&$(2,12)$\\
\strr & & $\IZ_2 \times \IZ_2$&$3$&$(3,13)$\\  \hline
\str$7487$&$(5,45)$&$\IZ_2$&$1-2$&$(5,25)$\\
\strr & & $\IZ_2 \times \IZ_2$&$3-6$&$(5,15)$\\  \hline
\str$7491$&$(4,44)$&$\IZ_2$&$1-4$&$(4,24)$\\
\strr & & $\IZ_2 \times \IZ_2$&$5-19$&$(4,14)$\\  \hline
\str$7522$&$(4,44)$&$\IZ_2$&$1$&$(4,24)$\\
\strr & & $\IZ_2 \times \IZ_2$&$2$&$(4,14)$\\  \hline
\str$7636$&$(3,47)$&$\IZ_2$&$1$&$(2,24)$\\  \hline
\str$7647$&$(3,47)$&$\IZ_2$&$1$&$(2,24)$\\  \hline
\str$7664$&$(5,50)$&$\IZ_3$&$1$&$(3,18)$\\  \hline
\str$7669$&$(3,48)$&$\IZ_3$&$1-2$&$(3,18)$\\
\strr & & \tr{$\IZ_3 \times \IZ_3$}&\tr{$3-8$}&\tr{$(3,8)$}\\  \hline
\str$7709$&$(6,54)$&$\IZ_2$&$1-2$&$(5,29)$\\  \hline
\str$7714$&$(3,51)$&$\IZ_2$&$1$&$(3,27)$\\
\strr & & $\IZ_2 \times \IZ_2$&$2$&$(3,15)$\\  \hline
\str$7719$&$(4,52)$&$\IZ_2$&$1$&$(4,28)$\\  \hline
\str$7731$&$(6,54)$&\tr{$\IZ_2$}&\tr{$1$}&\tr{$(5,29)$}\\  \hline
\str$7735$&$(3,51)$&$\IZ_2$&$1-2$&$(3,27)$\\
\strr & & $\IZ_4$&$3$&$(2,14)$\\
\strr & & $\IZ_2 \times \IZ_2$&$4-5$&$(3,15)$\\
\strr & & $\IZ_4 \times \IZ_2$&$6-8$&$(2,8)$\\  \hline
\str$7736$&$(4,52)$&$\IZ_2$&$1-3$&$(4,28)$\\  \hline
\str$7742$&$(4,52)$&$\IZ_2$&$1$&$(4,28)$\\  \hline
\str$7745$&$(3,51)$&$\IZ_2$&$1$&$(3,27)$\\
\strr & & $\IZ_4$&$2$&$(2,14)$\\
\strr & & $\IZ_2 \times \IZ_2$&$3$&$(3,15)$\\
\strr & & $\IZ_4 \times \IZ_2$&$4-5$&$(2,8)$\\  \hline
\str$7761$&$(2,52)$&$\IZ_2$&$1$&$(1,26)$\\
\strr & & $\IZ_5$&$2$&$(2,12)$\\
\strr & & $\IZ_{10}$&$3$&$(1,6)$\\  \hline
\str$7788$&$(3,55)$&$\IZ_2$&$1$&$(3,29)$\\  \hline
\str$7792$&$(3,55$)&$\IZ_2$&$1$&$(3,29)$\\  \hline
\str$7800$&$(5,59)$&$\IZ_3$&$1$&$(3,21)$\\  \hline
\str$7808$&$(2,56)$&$\IZ_3$&$1$&$(2,20)$\\
\strr & & $\IZ_3 \times \IZ_3$&$2-3$&$(2,8)$\\  \hline
\str$7810$&$(5,59)$&$\IZ_3$&$1-3$&$(3,21)$\\  \hline
\str$7819$&$(2,58)$&$\IZ_2$&$1$&$(2,30)$\\
\strr & & $\IZ_2 \times \IZ_2$&$2$&$(2,16)$\\  \hline
\str$7822$&$(2,58)$&$\IZ_2$&$1$&$(2,30)$\\  \hline
\str$7823$&$(2,58)$&$\IZ_2$&$1$&$(2,30)$\\
\strr & & $\IZ_2 \times \IZ_2$&$2$&$(2,16)$\\  \hline
\str$7861$&$(1,65)$&$\IZ_2$&$1$&$(1,33)$\\
\strr & & $\IZ_4$&$2$&$(1,17)$\\
\strr & & $\IZ_2 \times \IZ_2$&$3$&$(1,17)$\\
\strr & & $\IZ_8$&$4$&$(1,9)$\\
\strr & & $\IZ_4 \times \IZ_2$&$5-6$&$(1,9)$\\
\strr & & $\IQ_8$ & $7$ & $(1,9)$\\
\strr & & $\IZ_2 \times \IZ_2 \times \IZ_2$&$8$&$(1,9)$\\
\strr & & $\IZ_4 \times \IZ_4$&$9-11$&$(1,5)$\\
\strr & & $\IZ_4 \rtimes \IZ_4$ & $12$ & $(1,5)$\\
\strr & & $\IZ_8 \times \IZ_2$&$13$&$(1,5)$\\
\strr & & $\IZ_4 \times \IZ_2 \times \IZ_2$&$14-16$&$(1,5)$\\
\strr & & $\IZ_2 \times \IQ_8$ & $17-19$ & $(1,5)$\\
\strr & & $(\IZ_4 \times \IZ_2) \rtimes \IZ_4$ & $20-23$ & $(1,3)$\\ 
\strr & & $\IZ_8 \times \IZ_4$&$24-25$&$(1,3)$\\
\strr & & $\IZ_8 \rtimes \IZ_4$ & $26$ & $(1,3)$\\
\strr & & $(\IZ_8 \times \IZ_2) \rtimes \IZ_2$ & $27$ & $(1,3)$\\ 
\strr & & $\IZ_8 \rtimes \IZ_4$ & $28$ & $(1,3)$\\
\strr & & $\IZ_4 \times \IZ_4 \times \IZ_2$&$29-36$&$(1,3)$\\  
\strr & & $\IZ_2 \times (\IZ_4 \rtimes \IZ_4)$ & $37$ & $(1,3)$\\
\strr & & $\IZ_2 \times (\IZ_4 \rtimes \IZ_4)$ & $38$ & $(1,3)$\\
\strr & & $\IZ_4 \rtimes \IQ_8$ & $39$ & $(1,3)$\\
\strr & & $\IZ_2 \times \IZ_2 \times \IQ_8$ & $40-45$ & $(1,3)$\\ \hline
\str$7862$&$(4,68)$&$\IZ_2$&$1$&$(4,36)$\\
\strr & & $\IZ_4$&$2$&$(2,18)$\\
\strr & & $\IZ_2 \times \IZ_2$&$3$&$(4,20)$\\
\strr & & $\IZ_8$&$4$&$(1,9)$\\
\strr & & $\IZ_4 \times \IZ_2$&$5$&$(2,10)$\\
\strr & & $\IQ_8$ & $6$ & $(1,9)$\\
\strr & & $\IZ_4 \times \IZ_4$&$7$&$(1,5)$\\
\strr & & $\IZ_4 \rtimes \IZ_4$ & $8$ & $(1,5)$\\
\strr & & $\IZ_8 \times \IZ_2$&$9$&$(1,5)$\\  
\strr & & $\IZ_8 \rtimes \IZ_2$ & $10$ & $(1,5)$\\
\strr & & $\IZ_2 \times \IQ_8$ & $11$ & $(1,5)$\\ \hline
\str$7878$&$(1,73)$&$\IZ_3$&$1$&$(1,25)$\\
\strr & & $\IZ_3 \times \IZ_3$&$2-3$&$(1,9)$\\  \hline
\str$7884$&$(2,83)$&$\IZ_3$&$1$&$(2,29)$\\
\strr & & $\IZ_3 \times \IZ_3$&$2-5$&$(2,11)$\\  \hline
\str $7890$ & $(1,101)$ & $\IZ_5$ & 1 & $(1,21)$\\ 
\strr &   & $\IZ_5 \times \IZ_5$ & $2-5$ & $(1,5)$\\ \hline

\end{longtable}
\end{center}

\newpage
\bibliography{bibfile}{}
\bibliographystyle{utcaps}

\end{document}